\documentclass[a4paper,11pt,nofootinbib]{revtex4}

\usepackage{amsmath,amssymb,verbatim}
\usepackage{graphicx}
\usepackage{dcolumn}
\usepackage{bm}
\usepackage{xcolor}
\usepackage[all,color]{xy}
\usepackage{empheq}

\usepackage{tikz}

\usepackage{hyperref}

\newtheorem{theorem}{Theorem}
\renewcommand{\c}{\mathrm{c}}
\renewcommand{\b}{\mathrm{b}}
\newcommand{\mbar}{\br{\mc{M}}}
\newcommand{\br}[1]{\overline{#1}}
\newcommand{\zbar}{\overline{z}}
\renewcommand{\d}{\mathrm{d}}
\newcommand{\what}{\widehat}
\newcommand{\op}{\operatorname}
\newcommand{\C}{\mathbb{C}}

\newcommand{\R}{\mathbb{R}}

\newcommand{\Z}{\mathbb{Z}}

\newcommand{\eps}{\epsilon}
\newcommand{\g}{\mathfrak{g}}

\newcommand{\Oo}{\mc{O}}

\newcommand{\abracket}[1]{\left\langle#1\right\rangle}
\newcommand{\bbracket}[1]{\left[#1\right]}
\newcommand{\fbracket}[1]{\left\{#1\right\}}
\newcommand{\bracket}[1]{\left(#1\right)}

\newcommand{\mbb}{\mathbb}
\newcommand{\mf}{\mathfrak}
\newcommand{\mc}{\mathcal}

\newcommand{\pa}{\partial}

\newcommand{\dbar}{\bar\pa}

\newcommand{\iso}{\cong}

\newcommand{\A}{\mathcal A}

\renewcommand{\v}{\mathrm{v}}

\DeclareMathOperator{\Tr}{Tr}
\DeclareMathOperator{\PV}{PV}
\DeclareMathOperator{\End}{End}
\DeclareMathOperator{\HCS}{HCS}

\DeclareMathOperator{\Sym}{Sym}

\DeclareMathOperator{\BRST}{BRST}
\DeclareMathOperator{\KS}{KS}
\DeclareMathOperator{\BCOV}{BCOV}

\DeclareMathOperator{\ad}{ad}

\newcommand{\gl}{\mf{gl}}

\begin{document}


\title{Anomaly cancellation in the topological string}

\author{Kevin Costello}
 \email{kcostello@perimeterinstitute.ca}
\affiliation{%
 Perimeter Institute of Theoretical Physics\\
31 Caroline St N, Waterloo, ON N2L 2Y5, Canada
}%


\author{Si Li}
 \email{sili@mail.tsinghua.edu.cn}
\affiliation{
 Department of Mathematical Sciences and Yau Mathematical Sciences Center, \\
 Jingzhai, Tsinghua University, Beijing 100084, China
}%


\date{\today}

\begin{abstract}
We describe the coupling of holomorphic Chern-Simons theory at large N with Kodaira-Spencer gravity. We explain a new anomaly cancellation mechanism at all loops in perturbation theory for open-closed topological B-model. At one loop this anomaly cancellation is analogous to the Green-Schwarz mechanism.  
	
	As an application, we introduce a type I version of Kodaira-Spencer theory in complex dimensions $3$ and $5$. In complex dimension $5$, we show that it can only be coupled consistently at the quantum level to holomorphic Chern-Simons theory with gauge group $SO(32)$. This is analogous to the Green-Schwarz mechanism for the physical type I string. This coupled system is conjectured to be a supersymmetric localization of type I string theory. In complex dimension $3$, the required gauge group is $SO(8)$.  
\end{abstract}

\maketitle

\tableofcontents

\section{\label{sec:intro}Introduction}

Holomorphic Chern-Simons theory  is the open-string field theory of the $B$-model topological string \cite{W}. The corresponding closed-string field theory is known as Kodaira-Spencer theory \cite{BCOV}. This is a ``gravitational'' theory describing fluctuations of the complex structure of a Calabi-Yau manifold.  Both theories are non-renormalizable by power-counting, and so in principle could have new counter-terms or gauge anomalies appearing at arbitrary order in the loop expansion.  The problem of quantizing Kodaira-Spencer theory can be viewed as a toy model for the fundamental problem of quantizing Einstein gravity. Kodaira-Spencer theory, like Einstein gravity, is a non-renormalizable theory controlling geometric deformations of the space-time.  

String theory provides a mechanism for quantizing certain supergravity theories, due to the absence  of UV divergences.  Similarly, one can hope that topological string theory could provide a mechanism for quantizing Kodaira-Spencer theory.

In the first part of this paper we report on our work \cite{CL, CL2}, where we found a remarkable new anomaly-cancellation mechanism which allows us to quantize Kodaira-Spencer theory coupled to holomorphic Chern-Simons theory.  Our mechanism \emph{does not} require world-sheet techniques, only standard space-time Feynman diagrams.  Our mechanism cancels all anomalies occurring at all orders in the loop expansion, and fixes all counter-terms uniquely. At one loop the anomaly cancellation is analogous to the Green-Schwarz mechanism \cite{GS}.  

We view our method as a space-time proof of the UV finiteness of string theory.   

One reason a physicist might be interested in Kodaira-Spencer theory is that, on a Calabi-Yau manifold of dimension $5$, Kodaira-Spencer theory conjecturally describes \cite{ Berkovits:2003pq, CL2, CL3} a supersymmetric sector of type IIB superstring theory. To further test our method, we introduce in this paper a type I version of Kodaira-Spencer theory which lives on Calabi-Yau manifolds of complex dimension $3$ or $5$.  Type I Kodaira-Spencer theory on $\C^5$ can be coupled to $\op{SO}(n)$ holomorphic Chern-Simons theory, and a one-loop anomaly can be cancelled only if $n = 32$, as in the usual Green-Schwarz mechanism.  In that case, and only that case, our method also cancels all higher-loop anomalies.  This result is in some ways stronger than the original one of Green-Schwarz, because it holds to all orders in perturbation theory and not just at leading order. It would be very interesting to see whether our higher-loop anomaly cancellation mechanism can be applied to physical theories.

We conjecture that our type I Kodaira-Spencer theory, coupled to $\op{SO}(32)$ holomorphic Chern-Simons theory, is a supersymmetric localization of type I string theory.  This conjecture is further supported by a calculation where we show that the theory on a $D1$ brane in our type I theory matches a supersymmetric localization of the worldsheet theory of the $\op{Spin}(32)/\Z_2$ heterotic string.  

On $\C^3$, we show that the Green-Schwarz mechanism cancels the anomaly for type I Kodaira-Spencer theory when the holomorphic Chern-Simons gauge group is $SO(8)$.  We leave detailed investigation of this type I topological string to future work.  

\section{Open-closed topological B-model}
\subsection{The open-string sector}
In this section we will describe the open-string sector of toplogical $B$-model via Holomorphic Chern-Simons theory \cite{W}. Let us start with $X=\C^3$ the three dimensional complex space with linear holomorphic coordinates $\{z^i\}_{i=1,2,3}$. We denote 
$
\Omega_X^{p,q}
$ or simply $\Omega^{p,q}$ to be smooth differential forms of type $(p,q)$. An element $\alpha$ of $\Omega_X^{p,q}$ can be written as 
$$
\alpha= \sum_{i_1, \cdots, i_p, \bar j_1, \cdots, \bar j_q} \alpha_{i_1\cdots i_p \bar j_1\cdots \bar j_q} dz^{i_1}\wedge \cdots \wedge dz^{i_p}\wedge d\bar z^{\bar j_1}\cdots d \bar z^{\bar j_q}
$$
where $ \alpha_{i_1\cdots i_p \bar j_1\cdots \bar j_q}$'s are smooth functions on $X$ and they are totally skew-symmetric with respect to $i$-indices and with respect to $\bar j$-indices.

 Let $\g$ be a Lie algebra with non-degenerate Killing pairing $\Tr $. The fundamental field of holomorphic Chern-Simons theory  is a $\g$-valued $(0,1)$-form
 $$
A\in  \Omega_X^{0,1}\otimes \g. 
 $$
The  holomorphic Chern-Simons functional is given by 
\begin{equation}
S[A]:= \int_X \Tr\bracket{{1\over 2} A\wedge \dbar A+{1\over 6}A\wedge [A, A]}\wedge \Omega_X 
\end{equation}
where
\begin{equation}
\Omega_X= dz^1\wedge dz^2\wedge dz^3
\end{equation}
is the canonical holomorphic volume form. The bracket operation $[-,-]$ is induced by the Lie bracket on the $\g$-factor combined with wedge product on the form factor.  

The equation for $A$, obtained from varying $S$, is 
\begin{equation}
  \dbar A+{1\over 2}[A, A]=0. 
\end{equation}
This is equivalent to saying that $\dbar +A$ defines a new $(0,1)$-connection whose curvature vanishes. Geometrically, such data describes a holomorphic structure on the associated complex vector bundles with $\g$-action, where $\dbar +A$ describes the new $\dbar$-operator. 

The holomorphic Chern-Simons functional has a gauge symmetry. The infinitesimal gauge transformation is 
\begin{equation}
\delta_\phi A=\dbar \phi+ [A, \phi], \quad \phi \in \Omega_X^{0,0}\otimes \g.
\end{equation}
It is easy to verify that $\delta_\phi S=0$ for any such $\phi$. Solutions of the linearized equations of motion modulo gauge transformations are parametrized by the sheaf cohomology $H^1(X, \g)$ which is zero on the affine space $X=\C^3$. This corresponds to the fact that any holomorphic vector bundle on $\C^3$ is equivalent to a trivial holomorphic bundle. 

To obtain non-trivial solutions, we can put holomorphic Chern-Simons theory on an arbitrary complex three-dimensional manifold $X$ with a holomorphic volume form $\Omega_X$. Such a pair $(X, \Omega_X)$ is called a Calabi-Yau 3-fold.  $\g$ can be replaced by the endomorphism bundle $\End(E)$ of a holomorphic vector bundle $E$ on $X$, or the adjoint bundle associated to a holomorphic principal bundle. Then the moduli space of classical solutions modulo gauge transformations is given by all holomorphic structures on $E$. The infinitesimal deformation is again described by the cohomology $H^1(X, \End(E))$ which is non-zero in general. 

In this article, we will focus on the local case $X=\C^3$ to illustrate the basic idea of its coupling with gravity in the large N limit. 

To incorporate gauge fixing, we work with the Batalin-Vilkovisky (BV) formalism \cite{BV}.  We add into the ghost field living in $\Omega^{0,0}\otimes \g$, the anti-field of $A$ living in $\Omega^{0,2}\otimes \g$, and the anti-field of the ghost living in $\Omega^{0,3}\otimes \g$. The master field collecting all above is described by 
\begin{equation}
\A\in \Omega^{0,\bullet}\otimes \g [1]. 
\end{equation}
Here $[1]$ means a degree shifting on the space $\Omega^{0,\bullet}\otimes \g$ such that fields in $\Omega^{0,p}\otimes \g$ have cohomological degree  $p-1$ (ghost number $1-p$). We will sometimes talk about ``cohomological degree" in this paper to be consistent with homological algebra conventions latter. The space $\Omega^{0,\bullet}\otimes \g [1]$ has an odd symplectic structure of degree $-1$. The symplectic pairing $(-,-)$ is given by 
\begin{equation}
     (\eta_1, \eta_2)= \int_X \Tr(\eta_1\wedge \eta_2)\wedge \Omega_X, \quad \text{for}\quad \eta_1\in \Omega^{0,p}\otimes \g, \ \eta_2\in \Omega^{0,3-p}\otimes \g.
\end{equation}

The odd symplectic pairing $(-,-)$ induces a BV anti-bracket. Explicitly, let us choose a basis $e_\alpha$ of $\g$. Let $\eta_{\alpha\beta}=\Tr(e_\alpha e_\beta)$ and $\eta^{\alpha\beta}$ be the inverse matrix of $\eta_{\alpha\beta}$. We write $\A$ in components 
$$
\A=\sum_{\alpha}\A^\alpha e_\alpha, \quad  \A^\alpha=\sum_{p}\A_p^\alpha e_\alpha, \quad \A_p^\alpha \in \Omega^{0,p}.
$$
Then the BV anti-bracket is of the form
\begin{equation}
  \{\A(z)^\alpha, \A(w)^\beta\}_o=\eta^{\alpha\beta}\delta(z-w) \prod_{i=1}^3 (d\bar z^i-d\bar w^i)
\end{equation}
where $\delta(z-w)$ is the $\delta$-function on $\C^3$. The subscript $o$ refers to open string sector. The above formula is read by matching both sides with appropriate form components in $d\bar z_i$ and $d\bar w_i$. For example, the BV bracket is only nontrivial between $\A^\alpha_p$ and $\A^\beta_{3-p}$. The BV completed action of holomorphic Chern-Simons takes the same form 
\begin{equation}
\HCS[\A]:= \int_X \Tr\bracket{{1\over 2} \A\wedge \dbar \A+{1\over 6}\A\wedge [\A, \A]}\wedge \Omega_X
\end{equation}
where now $\A\in \Omega^{0,\bullet}\otimes \g [1]$ collects all fields of different types via its components. This is completely similar to the superspace formalism of ordinary Chern-Simons theory in the BV set-up \cite{AS}.

The following classical master equation 
\begin{equation}
\{\HCS, \HCS\}_o=0   \label{CME-HCS}
\end{equation}
holds, where $\{-,-\}_o$ is the above BV anti-bracket. This equation is equivalent to the fact that $\Omega^{0,\bullet}\otimes \g$ forms a differential graded Lie algebra, with $\dbar$ the differential and the Lie bracket induced from that on $\g$. We will denote the BRST operator $\delta_{\HCS}=\{\HCS,-\}$. Explicitly 
\begin{equation}
\delta_{\HCS}(\A)=\dbar \A+{1\over 2}[\A, \A].  \label{BRST-HCS}
\end{equation}
The classical master equation implies $\delta_{\HCS}^2=0$ as usual. 

\subsection{The closed-string sector}
In this section we will describe the closed-string sector of the topological $B$-model via Kodaira-Spencer gravity \cite{BCOV}, which  is generalized in \cite{CL} by turning on gravitational descendants. Let 
\begin{equation}
  \PV^{\bullet, \bullet}_X =\bigoplus\limits_{i,j}\PV^{i,j}_X
\end{equation}
denote the space of polyvector fields on a Calabi-Yau three-fold $X$. Here 
$
\PV^{i,j}
$ are smooth sections of the bundle $\wedge^i T^{1,0}_X\otimes \wedge^j (T_X^{0,1})^*$. In coordinates $\{z^i\}$, $\mu\in\PV^{i,j}$ has the form 
$$
\mu= \sum_{i_1, \cdots, i_p, \bar j_1, \cdots, \bar j_q} \mu^{i_1\cdots i_p}_{\bar j_1\cdots \bar j_q} \pa_{z^{i_1}}\wedge \cdots \wedge \pa_{z^{i_p}}\wedge d\bar z^{\bar j_1}\cdots d \bar z^{\bar j_q}
$$
where $ \mu^{i_1\cdots i_p}_{\bar j_1\cdots \bar j_q}$'s are smooth functions on $X$ and they are totally skew-symmetric with respect to $i$-indices and with respect to $\bar j$-indices.  $\pa_{z^i}$ is the holomorphic $(1,0)$-vector along $z^i$. We will write $|\mu|=i+j$ for the total degree of $\mu \in \PV^{i,j}$. 

We can identify $\PV^{i,j}$ with $\Omega^{3-i,j}$ via contracting polyvectors  with the $(3,0)$ volume form $\Omega_X$
\begin{equation}
  \PV^{\bullet, \bullet}\stackrel{\vdash \Omega_X}{\to} \Omega^{3-\bullet, \bullet}, \quad     \mu \to \mu\vdash \Omega_X. 
\end{equation}
The two operators $\dbar, \pa$ on differential forms $\Omega^{\bullet, \bullet}$ induce two linear operators on $\PV^{\bullet, \bullet}$, which we denote by $\dbar, \pa_{\Omega}$. 
Geometrically, 
\begin{equation}
\pa_\Omega: \PV^{p, q}\to \PV^{p-1, q}
\end{equation}
represents the holomorphic divergence operator with respect to the Calabi-Yau volume form $\Omega_X$. 

In the topological $B$-model, the complex $\PV^{\bullet,\bullet}$, with the differential $\dbar$, is the space of local operators on the world-sheet.  As in the physical string, not all local operators of the world-sheet topological field theory (TFT) can be modifications of the closed-string background: we should only consider operators which are invariant under world-sheet reparametrizations.  Because the world-sheet theory is a TFT, a small reparametrization will act trivially (up to a cochain homotopy) on the space of local operators.  Large reparametrizations can have non-trivial effects, however.  

Let us consider how this works in a general two-dimensional oriented TFT, with cochain complex of local operators $(V,Q_V)$.  We let $\op{Diff}_0(D)$ be the group of orientation-preserving reparametrizations of a small disc, with Lie algebra $\op{Vect}_0(D)$.  Then, $V$ has an action of $\op{Diff}_0(D)$ which is homotopically trivialized at the level of the Lie algebra $\op{Vect}_0(D)$.  

This is the kind of structure that is familiar from equivariant cohomology. Suppose $M$ is a manifold with an action of a Lie group $G$. The complex $\Omega^{\bullet}_M$ of differential forms on $M$ has a $G$-action, which at the level of the Lie algebra $\g$ is given by Lie derivative $\mc{L}_Y$ for $Y\in \g$.  The Cartan homotopy formula tells us that
\begin{equation*} 
	\mc{L}_Y = [\d_{dR}, \iota_Y] 
\end{equation*}
where $\iota_Y$ is the operator of contraction with the vector field generated by $Y$; $d_{dR}$ is the de Rham differential. That is, every $\mc{L}_Y$ acts homotopically trivially on $\Omega^{\bullet}_M$, so that we have a homotopy trivialization of the $\g$-action on $\Omega^{\bullet}_M$. In this situation, we know that the correct notion of $G$-invariants is given by the equivariant de Rham complex of $M$, defined to be the $G$-invariants in $\Omega^\bullet_M \otimes \Sym^\ast (\g^\vee)$, equipped with a differential combining the de Rham operator and $\iota_Y$. 

Similarly, for a TFT, invariance under world-sheet reparametrization is imposed by taking the equivariant cohomology with respect to the group $\op{Diff}_0(D)$. Since $\op{Diff}_0(D)$ is homotopy equivalent to the group $U(1)$ (rotations on $D$), we can just as well take the equivariant cohomology with respect to the $U(1)$ action.   The $U(1)$ action on the cochain complex $V$ of local operators of the world-sheet TFT is realized by a linear operator $\rho : V \to V$ which commutes with the BRST operator on $V$ and has integer eigenvalues.  The fact that small variations of a reparametrization are equivalent tells us that we must have an operator $D : V \to V$ with the feature that
\begin{equation}
	\rho = [D , Q_V]
\end{equation}
where $Q_V$ is the BRST operator on $V$. The $U(1)$-equivariant cohomology of $V$ is defined as follows. We first consider the subspace $V^\rho$ of operators invariant under $\rho$, then introduce an equivariant parameter $u$ of cohomological degree $2$, to give us the space $V^\rho[[u]]$, and use the differential $Q_V + u D$.  Here we work with formal series in $u$ instead of polynomials. 

In the context of the topological $B$-model, where $V = \PV_X^{\bullet,\bullet}$ and $Q_V=\dbar$ \cite{W2}, one finds by a TFT analysis \cite{Kontsevich:2006jb,Costello:2004ei, Lurie:2009keu}\footnote{These references identify the circle-rotation operator $D$ with the Connes $B$-operator on Hochschild cochains. By results of Connes \cite{Connes:1985} this is equivalent to the operator $\partial_{\Omega}$ on polyvector fields.}  that $\rho = 0$ and $D = \partial_\Omega$. Thus, the space of closed-string states -- that is, the $S^1$ equivariant cochains of  local operators of the worldsheet TFT -- is 
\begin{equation}
	\PV^{\bullet,\bullet}_X[[u]] \quad \text{with differential}\quad \dbar + u \partial_\Omega.
\end{equation}
This is the space of closed string fields of topological $B$-model proposed in \cite{CL}, as the enlargement of Kodaira-Spencer gravity \cite{BCOV} to include fields corresponding to world-sheet descendents. This complex played a fundamental role in the Barannikov-Kontsevich construction \cite{BK} of flat structures associated to the Kodaira-Spencer gravity.  

\subsection{Coupling closed to open strings at leading order}\label{section-coupliing}
Every closed-string state $\Sigma \in \PV^{\bullet,\bullet}[[u]]$ will give us a first-order deformation $I^{(1)}_{\Sigma}$ of the holomorphic Chern-Simons action.  This can be understood from a Lie algebra cohomology computation on the BRST complex of  holomorphic Chern-Simons theory. This was described in \cite{CL2}, where we gave an explicit formula of the deformation (including also higher order deformations at the disk level)  from a world-sheet calculation of disc amplitudes \cite{WC} (extending that of \cite{K}).   

It is important to emphasize that writing down disc amplitudes at all orders in the open and closed string fields is a \emph{very} non-trivial problem. Indeed, giving a collection of such disc amplitudes which solve the master equation is equivalent to providing a proof of (the cyclic enhancement of) Kontsevich's formality theorem, proved in \cite{WC, K}.  

For instance, we note that any holomorphic and divergence-free Poisson tensor on the space-time $\C^n$ satisfies the closed-string equations of motion. If we have a collection of disc amplitudes, we can insert this Poisson tensor as a closed string field.  The master equation for the disc amplitudes implies that the open-string amplitudes in the presence of this closed-string field satisfy the axioms of an $A_\infty$ algebra. If we have a space-filling brane on $\C^n$, the cohomology of this $A_\infty$ algebra is some deformation of the associative algebra $\C[z_1,\dots,z_n]$. We conclude, that providing a collection of disc amplitudes for the open-closed theory gives a universal formula for deformation quantizing any divergence-free holomorphic Poisson tensor.  There is no simple expression for such a universal formula: the problem of providing one was solved by Kontsevich \cite{K} by explicitly evaluating the world-sheet path integral for the disc amplitudes. 

To get a feeling for the deformation $I^{(1)}_{\Sigma}$,  we start with the case when $\Sigma=\mu\in \PV^{k, \bullet}$ has no higher $u$-order and $\pa_\Omega \mu=0$, i.e. $\mu$ is divergence free.  There $I^{(1)}_\mu$ is easy to describe and is proportional to 
\begin{equation} 
	I^{(1)}_\mu[\A]\propto \sum_{i_1, \cdots, i_k}  \int_{X} \Tr  \bracket{\mu^{i_1\cdots i_k} \A\wedge \pa_{z^{i_1}} \A\wedge\cdots \wedge \pa_{z^{i_k}}\A}\wedge \Omega_X.  \label{1st-order-coupling}
\end{equation}
Here $\A\in \Omega^{0, \bullet}\otimes \gl_N[1]$ is the master field of HCS, and $\mu=\sum_{i_1\cdots i_k}\mu^{i_1\cdots i_k} \pa_{z^{i_1}}\wedge \cdots \wedge \pa_{z^{i_k}}$ where $\mu^{i_1\cdots i_k}\in \Omega^{0,\bullet}$. We can write the above formula in a compact notation as
$$
\int_{X} \Tr  \bracket{\mu\vdash \A\wedge \pa \A\wedge\cdots \wedge \pa \A}\wedge \Omega_X.
$$


For instance, if $\mu=\sum_{\bar i, j}\mu_{\br{i}}^{j} d \bar z^{\bar i} \pa_{z^j}\in \PV^{1,1}$, then this expression describes the response of the holomorphic Chern-Simons action to a fluctuation in the complex structure of $X$:
\begin{equation} 
	I^{(1)}_{\mu}[\A] \propto \sum\limits_{\bar i, j}\int_{X} \mu_{\br{i}}^j \Tr \d \zbar^{\br{i}}\bracket{ \A \wedge \pa_{z^j} \A} \wedge \Omega_X.
\end{equation}
This term has the effect of varying $\dbar$ to $\dbar + \sum_{\bar i, j}\mu_{\br{i}}^j \d \zbar^{\br{i}} {\pa\over \partial {z^j}}$ in the holomorphic Chern-Simons action. 

If $\mu=\sum_{i,j}\mu^{ij}\pa_{z^i}\wedge \pa_{z^j} \in \PV^{2,0}$, then $I^{(1)}_{\mu}$ gives the infinitesimal response of the holomorphic Chern-Simons action to making the space-time non-commutative, so that the coordinates commute to leading order as  $[z^i, z^j] = \mu^{ij}(z)$.
In this case, $I^{(1)}_{\mu}$ is
\begin{equation} 
	I^{(1)}_\mu[\A]\propto \sum_{i,j}\int_{X} \Tr  \bracket{\mu^{ij} \A\wedge \pa_{z^{i}} \A\wedge \pa_{z^{j}}\A}\wedge \Omega_X.  
\end{equation}
This term arises from $\Tr \bracket{\A \wedge \A \wedge \A}$ when the space-time coordinates fail to commute.

The coupling of a general element of $\PV^{\bullet,\bullet}[[u]]$ is a little more complicated. If we take an element $u^l \mu$ for $\mu \in \PV^{k,\bullet}$, then the coupling $I^{(1)}_{u^l \mu}$ is a sum of terms of the form
\begin{equation} 
	\int_{X} \Tr  \bracket{\mu^{i_1\cdots i_k} \A\wedge \A^{r_1} \wedge \pa_{z^{i_1}} \A\wedge   \cdots \wedge \A^{r_k} \wedge \pa_{z^{i_k}}\A}\wedge \Omega_X 
\end{equation}
where $r_1 + \dots r_k = 2l$. Here  $\A^{m}$ means the $m$-th wedge product $\A\wedge\cdots \wedge \A$. That is, the coupling for $u^l \mu$ takes the same form as the coupling for $\mu$, except that we place extra $2l$ copies of $\A$ without any $z$-derivatives at general places in the cyclic word we are integrating, with certain extra coefficients. 

The simplest example of this is when $\Sigma=u\mu$ for $\mu\in \PV^{0,0}$, in which case 
\begin{equation} 
	I^{(1)}_{u \mu} \propto \int_X \mu \Omega_X \wedge \op{Tr} (\A \wedge \A \wedge \A). 
\end{equation}
This term changes the interacting term of the holomorphic Chern-Simons action. If $\dbar \mu = 0$, then a field redefinition of the form $\A \to \A + \mu \A$ shows that $I^{(1)}_{u \mu}$ is the response of the holomorphic Chern-Simons action to a change in the volume form $\Omega_X \to \Omega_X (1 + \mu)$ (at leading order).

\subsection{Kodaira-Spencer gravity in BV formalism}\label{Sec-KS}

The space $\PV^{\bullet, \bullet}[[u]][2]$ describes gravitational modes in the B-twisted sector of topological string. Here $[2]$ is the degree shift by $2$ such that elements in $u^k \PV^{i,j}$ have degree $i+j+2k-2$ (ghost number $2-i-j-2k$). Such shift puts $\PV^{1,1}$ at degree $0$, which is the field describing deformation of complex structures on $X$. Following \cite{BCOV}, we still call it the Kodaira-Spencer gravity. 

$\PV^{\bullet, \bullet}$ has a natural differential BV structure, with the differential 
\begin{equation}
Q=\dbar+u\pa_\Omega
\end{equation}
and the odd bracket $[-,-]$ coming from the Schouten-Nijenhuis bracket. We normalize the sign convention such that the following BV relation holds
\begin{equation}\label{eqn-BV}
 [\alpha, \beta]=\pa_\Omega(\alpha\beta)-(\pa_\Omega \alpha)\beta-(-1)^{|\alpha|}\alpha \pa_\Omega \beta, \quad \forall \alpha, \beta\in \PV^{\bullet, \bullet}. 
\end{equation}
The equation of motion describing Kodaira-Spencer gravity is given by the Maurer-Cartan equation 
\begin{equation}
Q\Sigma +{1\over 2}[\Sigma, \Sigma]=0, \quad \Sigma \in \PV^{\bullet, \bullet}[[u]].
\end{equation}
The solution space of this set of equations modulo gauge equivalence is shown \cite{BK} to lead to a \emph{smooth} moduli, as long as the manifold $X$ is compact. This generalizes the classical theorem of Bogomolov-Tian-Todorov on the smoothness of Calabi-Yau moduli. 

To understand the geometric meaning of its solution, let us consider the sub-locus when 
\begin{equation}
\Sigma=\mu+ u \rho, \quad \mu \in \PV^{1,1}, \rho\in \PV^{0,0}. \label{eqn-KS-field}
\end{equation}
Note that by the degree shifting $[2]$, $\Sigma$ of the above form has degree $0$.  The equations of motion state that  
\begin{align}\label{eqn-deformation-pair}
  \begin{cases} \dbar \mu+{1\over 2}[\mu, \mu]=0. \\
   \dbar \rho+\pa_\Omega \mu+[\mu, \rho]=0.
   \end{cases}
\end{align}
The first equation describes an integrable deformation of complex structure via the standard Kodaira-Spencer theory. In the new complex structure specified by $\mu$, the smooth form $e^{\mu}\vdash \Omega_X$ will have type $(3,0)$, but may not be holomorphic. The second equation is equivalent to (using the first equation)
\begin{equation}
\d(e^\rho e^{\mu}\vdash \Omega_X)=0. 
\end{equation}
This says that $e^\rho$ rescales the new $(3,0)$-form $e^{\mu}\vdash \Omega_X$ into a closed form. It follows that $e^\rho e^{\mu}\vdash \Omega_X$ precisely describes a new holomorphic volume form in the complex structure $\mu$.  Therefore the equation of motion for $(\mu, \rho)$ describes the pair deformation  $(X, \Omega_X)$ of a complex structure together with a Calabi-Yau volume form. i.e.,  deformation of Calabi-Yau structures.  We can also include other components of the fields, and they can be viewed as extended  Calabi-Yau structure, including all possible non-commutative deformations, as well as all possible symmetries. 

In the original formulation of \cite{BCOV}, Kodaira-Spencer fields are described by divergence free polyvectors (i.e. polyvectors preserving the Calabi-Yau volume form)
$$
\ker \pa_\Omega \subset \PV^{\bullet, \bullet}.
$$
The equation of motion is
\begin{equation}
\dbar \mu+{1\over 2}[\mu, \mu]=0, \quad \mu \in \ker \pa_\Omega
\end{equation}
which describes the deformation of complex structures. Our model $\PV^{\bullet, \bullet}[[u]]$ can be viewed as an extension of \cite{BCOV} by turning on the gravitational descendants. It extends complex structure deformations to deformations of Calabi-Yau structures. As we have seen, the model including the paremeter $u$ is natural from the string-field theory perspective.  

To connect this to Einstein gravity, we need to borrow Yau's theorem on Calabi conjecture, which says that on compact K\"{a}hler manifold, solutions of Ricci flat metrics in a certain K\"{a}hler class are completely determined by the complex structure. This explains the name ``Kodaira-Spencer gravity''.

It is non-trivial to write down an action functional leading to the equations of motion described above.  The kinetic term in the action functional is not that well defined,  although the propagator and the linearized BRST operator are defined.  The interaction term is however defined, and can in principle be computed using world-sheet methods.  We will write down the formula for the interaction shortly.


It will be important for our analysis of quantization to put Kodaira-Spencer gravity into the BV formalism.  That is, we will introduce a BV anti-bracket on the space of Kodaira-Spencer fields. Precisely, let us write $\Sigma\in \PV^{\bullet, \bullet}[[u]]$ in components 
$$
\Sigma=\sum_{k\geq 0}\mu_{k}u^k,  \quad \mu_{k}=\sum_{i,j=0}^3  \mu^{(i,j)}_{k}, \quad \text{where}\quad \mu^{(i,j)}_{k}\in \PV^{i,j}. 
$$
The only nontrival BV anti-bracket is given by 
\begin{align}
\{\mu_{0}(z), \mu_{0}(w)\}_c=\sum_{k=1}^3 {\pa \over \pa {z^k}}\delta(z-w) \prod_{i=1, i\neq k}^3(\pa_{z^i}-\pa_{w^i}) \prod_{j=1}^3(d\bar z^{j}-d\bar w^{\bar j}). \label{KS-BV-bracket}
\end{align}
Here $\delta(z-w)$ is the $\delta$-function. The factor $\prod_{i=1, i\neq k}^3(\pa_{z^i}-\pa_{w^i})=
(-1)^k (\pa_{z^1}-\pa_{w^1})\wedge\cdots \widehat{(\pa_{z^k}-\pa_{z^k})}\wedge \cdots (\pa_{z^3}-\pa_{w^3}). 
$ The subscript $c$ refers to closed string sector. 

The above formula is read by matching both sides with appropriate components of polyvectors. For example, the BV bracket is only nontrivial between $\mu^{(i,j)}_{0}$ and $\mu^{(2-i,3-j)}_{0}$. The BV bracket involving $\mu_{k}$ for $k>0$ are all zero. In other words, only fields $\mu_{0}^{(i,j)}$ with $i\neq 3$ are dynamical. In particular, this BV bracket is highly degenerate.

Intrinsically, the Poisson kernel of the above BV bracket is given by the integral kernel of the divergence operator $\pa_\Omega$ on polyvector fields. To illustrate this, we give an explicit description of the induced BV anti-bracket on local functions. Let us denote for convenience 
\begin{equation}
\int_X^{\PV}: \PV^{\bullet, \bullet}\to \C, \quad \mu \to \int_X (\mu \vdash \Omega_X)\wedge \Omega_X. 
\end{equation}
This is an integration on polyvector fields against the holomorphic volume form. Note that it is only nonzero on $\PV^{3,3}$. 

Let $S=S[\Sigma]$ be a local functional of $\Sigma=\sum\limits_{k\geq 0}\mu_{k}u^k\in  \PV^{\bullet, \bullet}[[u]]$. We define the variation ${\delta S\over \delta \mu_k}$ by
\begin{equation}
   \delta S:= \sum_{k\geq 0}  \int_X^{\PV} \delta \mu_k\wedge  {\delta S\over \delta \mu_k}, \quad {\delta S\over \delta \mu_k}\in \PV^{\bullet, \bullet}. 
\end{equation}
Then for any two local functional $S_1, S_2$ of $\Sigma$, their induced BV anti-bracket from \eqref{KS-BV-bracket} is given by 
\begin{equation}
  \{S_1, S_2\}_c= \int_X^{\PV} \bracket{\delta S_1\over \delta \mu_0}\pa_{\Omega}\bracket{\delta S_2\over \delta \mu_0}. \label{KS-BV-action} 
\end{equation}
It is not hard to check that it satisfies the graded Jacobi-identity, defining a consistent BV anti-bracket. This fact also follows from a general abstract analysis in \cite{CL}.

Define the following local functional which we call BCOV interaction \cite{CL} (See also \cite{LSS} for an algebraic combinatorial model)
\begin{equation}
  I^{\BCOV}[\Sigma]=\int_X^{\PV} \abracket{e^{\Sigma}}_0=\sum_{n\geq 3}{1\over n!}\int_X^{\PV} \abracket{\Sigma^{\otimes n}}_0, \quad \Sigma\in \PV^{\bullet, \bullet}[[u]].
\end{equation}
Here $\abracket{-}_0$ means 
\begin{equation}
\abracket{u^{k_1}\alpha_1\otimes \cdots\otimes u^{k_m}\alpha_m}_0:=\binom{m-3}{k_1\cdots k_m}\alpha_1\wedge \cdots\wedge \alpha_m \quad \text{for}\quad \alpha_i\in \PV^{\bullet, \bullet}.  
\end{equation}
Note that the leading cubic term is precisely the Kodaira-Spencer interaction described in \cite{BCOV}. This interaction is natural from the string-field theory point of view since
\begin{equation*} 
	\binom{m-3}{k_1\cdots k_m}= \int_{\mbar_{0,m}} \psi_1^{k_1} \dots \psi_m^{k_m}.
\end{equation*}
One can show that $I^{\BCOV}$ satisfies the following classical master equation \cite{CL}
\begin{equation} 
	QI^{\BCOV} + \tfrac{1}{2}\{I^{\BCOV}, I^{\BCOV}\}_c = 0. 
\end{equation}
Here $Q$ refers to the linearized BRST transformation transformation $\Sigma\to Q\Sigma=(\dbar+u\pa_\Omega)\Sigma$. This allows us to define the non-linear BRST transformation
\begin{equation}
\delta_{\BCOV}=Q+\{I^{\BCOV},-\}_c.
\end{equation}
Classical master equation implies $\delta_{\BCOV}^2=0$, which is in fact a nontrivial identity to check. 
 Similar to \eqref{BRST-HCS}, the equations of motion of the string field theory with interaction $I^{\BCOV}$ are 
 \begin{equation}
 \delta_{\BCOV}(\Sigma) = 0, \quad \text{for}\quad \Sigma \in \PV^{\bullet,\bullet}[[u]].
\end{equation}

Let us compare these equations with that in the Kodaira-Spencer gravity described above. From the Kodaira-Spencer gravity, it is more natural to consider the following BRST operator 
\begin{equation}
\delta_{\KS}(\Sigma)=Q \Sigma+ {1\over 2}[\Sigma, \Sigma]. 
\end{equation}
$\delta_{KS}(\Sigma)=0$ is precisely the Maurer-Cartan equation described above. 

It can be checked that $\delta_{\BCOV}$ is equivalent to $\delta_{\KS}$ under the nonlinear transformation of fields
\begin{equation}
\Sigma \to \bbracket{u(e^{\Sigma/u}-1)}_+, \quad \Sigma\in \PV^{\bullet, \bullet}[[u]]. 
\end{equation}
Here $[-]_+$ means truncating to non-negative powers in $u$. We refer to \cite{CL} for details about this (see also \cite{L-review} for a review).  It follows that the equation $\delta_{\BCOV}(\Sigma)=0$ again describes deformation of Calabi-Yau structures. 

We can not further express $Q$ as a BV bracket. Again, this is because the kinetic term in the Lagrangian of the string-field theory is ill-defined. The classical master equation for  $I^{\BCOV}$ fits into the general form of closed string field dynamics as described in \cite{Z1}. 

\subsection{Green-Schwarz mechanism}\label{sec-GS}

It is standard to quantize classical gauge theories within BV formalism in terms of quantum master equation. The precise form of the quantum master equation for open-closed string fields can be organized in terms of topological types of bordered Riemann surfaces \cite{Z1,Z2}. 

In general, the problem of solving quantum master equation may be obstructed due to gauge anomalies. It turns out that there is a remarkable one-loop anomaly cancellation for holomorphic Chern-Simons theory coupled to Kodaira-Spencer theory due to the interplay between open and closed string sectors. This is a topological string version of  the Green-Schwarz mechanism \cite{GS}. 

The one-loop gauge anomaly of holomorphic Chern-Simons theory  can be computed in a standard way using a diagram with four vertices. In the BV formalism, this is analyzed in \cite{CL2}, which is further systematically developed in \cite{B-holomorphic}.  Explicitly, we let 
$\A\in \Omega^{0,\bullet}\otimes \g [1]$ be the master field collecting fields, ghost, etc. Let 
\begin{equation}
F_{\A}:= \d\A+{1\over 2}[\A, \A]
\end{equation}
be the curvature form. Then the one-loop gauge anomaly is proportional to the expression
\begin{equation}
\int_X \Tr_{\ad} \A (F_{\A})^3.
\end{equation}
Here $\Tr_{ad}$ means the trace in the adjoint representation of $\g$.  Let us restrict our master field to the more familiar fields 
$$
\A=A+\c, \quad A\in \Omega^{0,1}\otimes \g, \c\in \Omega^{0,0}\otimes \g
$$
where $A$ is the connection and $c$ is the ghost field. Then the anomaly is proportional to the familiar expression 
\begin{equation}
 \int_X \Tr_{\ad}(\c (F_{A})^3) .
\end{equation}
Because $A$ is a $(0,1)$-form, this is the same as $\int \op{Tr}_{\ad}(\c (\partial A)^3)$. 

As a consistency test, we check the Wess-Zumino consistency condition.  Let us denote
$$
F_{\A}=F_{\A}^{1,\bullet}+ F_{\A}^{0, \bullet}
$$
where 
\begin{equation}
  F_{\A}^{1,\bullet}=\pa \A, \quad F_{\A}^{0, \bullet}=\dbar \A+{1\over 2}[\A, \A].
\end{equation}
The Bianchi identity $dF_{\A}+[\A, F_{\A}]=0$ implies 
\begin{equation}
   \pa F_{\A}^{1,\bullet}=0,\quad \pa F_{\A}^{0, \bullet}+ \dbar F_{\A}^{1,\bullet}+[\A, F_{\A}^{1,\bullet}]=0, \quad \dbar F_{\A}^{0,\bullet}+[\A, F_{\A}^{0,\bullet}]=0.
\end{equation}
Using this above relation, we find
\begin{align}
  \delta_{\HCS}\Tr (F_{\A}^k)&=-\pa \Tr (F_{\A})^k\\
  \delta_{\HCS}\Tr (\A (F_{\A})^k)&=-\pa \Tr (\A (F_{\A})^k)+\Tr (F_{\A})^{k+1}. 
\end{align}
Then the following Wess-Zumino consistency holds
\begin{equation}
\delta_{\HCS}\int_X \Tr_{\ad}(\A (F_{\A})^3=\int_X \Tr_{\ad}(F_{\A})^4=0
\end{equation}
since the topological term  $\Tr(F_{\A})^4$ is a total derivative. 

Now we specialize to the case when $\g=\gl(N)$. Since the adjoint representation $\ad$ of $\gl(N)$ is  the tensor of the fundamental representation $fun$ and its dual, we have 
\begin{align}
 \int_X \Tr_{\ad}(\A (F_{\A})^3)= \int_X \Tr_{\ad}(\A (\pa \A)^3)\propto \sum_{i=0}^3\int_X (-1)^i \Tr_{fun} \A (\pa \A)^i \Tr_{fun} (\pa \A)^{3-i}.
\end{align}

On the other hand, there is a tree level contribution from Kodaira-Spencer field 
\begin{equation*} 
	\{I^{(1)}, I^{(1)} \}_c 
\end{equation*}
where $I^{(1)}$ is the first order coupling of a Kodaira-Spencer field to a holomorphic Chern-Simons field as described in Section \ref{section-coupliing}. The BV bracket of Kodaira-Spencer field \eqref{KS-BV-action} implies that only terms  in \eqref{1st-order-coupling} of $I^{(1)}$ will contribute to $\{I^{(1)}, I^{(1)} \}_c$ and we find
\begin{equation}
	\{I^{(1)}, I^{(1)} \}_c =\sum_{i=0}^2 (-1)^i \int_X \Tr_{fun} \A (\pa \A)^i \Tr_{fun} (\pa \A)^{3-i}.
\end{equation}
By suitably rescaling $I^{(1)}$, they cancel with that in the one-loop gauge anomaly and we end up with a total anomaly proportional to
\begin{equation} 
	\Tr_{fun} (1)  \int_X \Tr_{fun} (\A (F_{\A})^3). 
\end{equation}
   
When $\g=\gl(N)$, $\Tr_{fun} (1)=N$. Then we find that the total anomaly at leading order is $N \int_X \Tr_{fun} (\A (F_{\A})^3)$, or $N  \int_X \Tr_{fun}(c F(A)^3)$ when $\A=A+c$.   We can get around this anomaly by working with the super Lie algebra $\gl({N|N})$ instead. The formulae for coupling the $\gl({N \mid N})$ gauge theory to Kodaira-Spencer theory are the same as those for $\gl(N)$. Since $\Tr_{fun}(1)$ is the super trace on $\gl({N|N})$, which vanishes, there is no anomaly in this case. 

The Green-Schwarz mechanism only accounts for one-loop anomalies. It turns out that anomalies cancel at all orders in the loop expansion \cite{CL2}. We will sketch this argument later in this note.

\subsection{The holomorphic stress-energy tensors}\label{sec-stress-tensors}
Before we get to explaining the cancellation of higher loop anomalies, we will need to study the holomorphic stress-energy tensor of holomorphic Chern-Simons theory.  

Any Lorentz invariant field theory on $\R^n$ has a stress-energy tensor $T^{ij}$, which is a local operator in the theory. This tensor tells us how to couple the theory to a variation in the metric tensor.  If we vary the metric to $\delta_{ij} + g_{ij}$, then the Lagrangian $\mc{L}$ of the theory varies to first order as 
\begin{equation*} 
	\mc{L}\to \mc{L} + \sum_{i,j}\int_{x \in \R^n} g_{ij}(x) T^{ij}(x). 
\end{equation*}

In a similar way, any holomorphic theory on $\C^n$ has a stress-energy tensor which tells us how the theory responds to a variation of the complex structure. The holomorphic stress-energy tensor $T_{l \br{i}_1 \dots \br{i}_{n-1}}$ (anti-symmetric in the indices $\br{i}_k$) is characterized by the fact that if we vary the complex structure by the Beltrami differential $\mu_{\br{i}}^{j}$, then, to first order, the Lagrangian of the holomorphic theory varies as
\begin{equation} 
\mc{L}\to	\mc{L} + \sum_{l, \br{i}_1,\cdots, \br{i}_n}\int_{z \in \C^n}\eps^{\br{i}_1\dots \br{i}_{n}} \mu_{\br{i}_n}^l(z) T_{\br{i}_1 \dots \br{i}_{n-1} l}(z) \label{eqn-stress-tensor}
\end{equation}
Here $\eps^{\br{i}_1\dots \br{i}_{n}} $ is the totally antisymmetric tensor in anti-holomorphic indices. 

The holomorphic stress-energy tensor $T_{l \br{i}_1 \dots \br{i}_{n-1}}$ represents a $(1,n-1)$-form with coefficients in the algebra of local operators.  Let us denote it by $T^{(1,n-1)}$. Then the above variation \eqref{eqn-stress-tensor} is read in a compact form
\begin{equation}
	\mc{L}\to \mc{L} +  \int_{\C^n} T^{(1,n-1)}\wedge (\mu\vdash \Omega),  \quad \mu \in \PV^{1,1}, \quad \Omega=d z^1\wedge  \cdots \wedge dz^n. \label{eqn_variation} 
\end{equation}
Here $\mu\vdash \Omega$ is contracting $\mu$ with the holomorphic volume form $\Omega$ to get an $(n-1,n)$-form \footnote{If we did not want to introduce a holomorphic volume form on our complex manifold, we could treat $T^{(1,n-1)}$ as an operator-valued section of the canonical bundle tensored with $(1,n-1)$ forms.}.  We then wedge with the operator-valued $(1,n-1)$ form $T^{(1,n-1)}$, to get an $(n,n)$ form valued in local operators, i.e. a first-order variation of the Lagrangian. This we can integrate over $\C^n$.

In the holomophic world, however, there is no reason for $T^{(1,n-1)}$ to be BRST closed. This is because we do not expect to obtain a consistent deformation of the theory by coupling to a Beltrami differential which is not $\dbar$-closed.  Instead, we expect that $Q_{\BRST} T^{(1,n-1)}$ should be $\dbar$-exact:
\begin{equation} 
	Q_{\BRST}  T^{(1,n-1)} =  \dbar T^{(1,n-2)} 
	\end{equation}
for some tensor $T^{(1,n-2)} $ which is a $(1,n-2)$-form.  This is sufficient to imply that the expression \eqref{eqn-stress-tensor} is BRST closed as long as $\mu$ is $\dbar$-closed:
$$
Q_{\BRST}\int_{\C^n}   T^{(1,n-1)} \wedge (\mu \vdash \Omega) = \int_{\C^n} (\dbar T^{(1,n-2)}) \wedge (\mu \vdash \Omega)=(-1)^{n}\int_{\C^n}   T^{(1,n-2)}) \wedge(\dbar \mu \vdash \Omega)=0. 
$$
Note that in complex dimension $1$, any Beltrami differential defines an integrable complex structure, which is why we always have a BRST closed stress-energy tensor in that case. 

Similarly, $T^{(1,n-2)}$ is not BRST closed, but its BRST variation is $\dbar$-exact:
\begin{equation*} 
	Q_{\BRST}  T^{(1,n-2)}=  \dbar T^{(1,n-3)}.
\end{equation*}
Iterating this procedure, we find that there is a BRST closed stress-energy tensor $T^{(1,0)}$ of ghost number $n-1$ ($T^{(1,n-1)}$ has ghost number $0$).  This is the fundamental holomorphic stress energy tensor. The other tensors $T^{(1,\bullet)}$ are obtained from $T^{(1,0)}$ by descent.

Let us specialize the above discussion to dimension $3$, and consider the stress-energy tensor of holomorphic Chern-Simons theory.  Here, there are additional constraints  coming from the fact that the theory is defined only on manifolds equipped with a holomorphic volume form which could vary together with the change of complex structures.  Thus, we will have a second kind of stress-energy tensor which we call $T^{(0,3)}_1$ which describes the response to a variation of the holomorphic volume form.   

If we vary to leading order the the holomorphic volume form by $\Omega_X(1 + \rho)$ for some $\rho \in \PV^{0,0}$, then the response of the theory to these changes in the geometry is given by
\begin{equation*} 
	\mc{L}\to	\mc{L} + \int_X T^{0,3}_1 \rho \Omega_X 
\end{equation*}
If we simultaneously vary the volume form by $\rho$ and complex structure by some Beltrami differential $\mu \in \PV^{1,1}$, the Lagrangian is varied by
\begin{equation*} 
	\mc{L}\to	\mc{L} + \int_X T^{1,2}_0 \wedge (\mu \vdash \Omega_X) +  \int_X T^{0,3}_1 \rho \Omega_X 
\end{equation*}
We do not expect every variation of complex structure and of volume form to give rise to a consistent deformation of the theory.   A BRST closed deformation of the Lagrangian (at leading order) should only arise when $\dbar \mu= 0$ and $\pa_\Omega \mu + \dbar \rho = 0$  (the linear part of \eqref{eqn-deformation-pair}). 

 These equations tell us that the two stress energy tensors $T_0^{(1,2)}$ and $T^{(0,3)}_1$ should satisfy 
\begin{align*} 
	Q_{\BRST}T_0^{(1,2)}&=\dbar T_0^{(1,1)}+ \pa T_1^{(0,2)} \\ 
	Q_{\BRST} T^{(0,3)}_1&= \dbar T^{(0,2)}_1. 
\end{align*}
Further iterating this procedure, we end up with the operator $T_1^{(0,0)}$ of ghost number $3$ and the operator $T_0^{(1,0)}$ of ghost number $2$, related by
\begin{align*} 
	Q_{\BRST} T_0^{(1,0)} &= \partial T_1^{(0,0)}\\
	Q_{\BRST} T^{(0,0)}_1 &= 0.
\end{align*}
This describes the stress-energy tensors we find for a holomorhpic theory that can be placed on a Calabi-Yau $3$-fold.

In holomorphic Chern-Simons theory, the stress-energy tensors $T^{(1,0)}_0$ is found from the first order coupling to a background field of Kodaira-Spencer theory in $\PV^{1,3}$. Equation \eqref{1st-order-coupling} tells us
\begin{equation} 
	T_0^{(1,0)}= \op{Tr} (\c \partial \c)  
\end{equation}
where $\c$ is the ghost field for the gauge transformations of holomorphic Chern-Simons theory (the ghost field is the lowest degree component of the master field $\A$). 

Similarly, the stress-energy tensor $T_1^{(0,0)}$ is given by coupling to the Kodaira-Spenser field in $\rho \in u \PV^{0,3}$, giving us 
\begin{equation} 
	T_1^{(0,0)} = \op{Tr} (\c^3). 
\end{equation}
Since $Q_{\BRST} \c = \c^2$ as usual, one can calculate readily that $Q_{\BRST} T_1^{(1,0)}=\partial  T_1^{(0,0)}$ holds.	

For a theory like holomorphic Chern-Simons for $\mf{gl}_N$, we can couple to an arbitrary Kodaira-Spencer field. Therefore we will have a tower of stress-energy tensors $T^{(k,m)}_l\in \Omega^{k,m}$ measuring the response to a variation by $ \mu^{(k,3-m)} u^l \in \PV^{k,3-m}u^l$ such that the Lagrangian changes at the leading order by 
\begin{equation}
\int_X T^{(k,m)}_l  \wedge \bracket{\mu^{(k,3-m)}\vdash \Omega_X}.
\end{equation}
The above discussion generalizes to all $T^{(k,m)}_l$'s ($l\geq 0$) and we find 
\begin{align}\label{stress-BRST-variation}
  \begin{cases}  Q_{BRST} T_l^{(k,m)} = \dbar T_l^{(k,m-1)}+\pa T_{l+1}^{(k-1,m)} & k,m\geq 1\\
        Q_{BSRT} T^{(0,m)}_l =\dbar T_l^{(0,m-1)} & m\geq 1\\
	Q_{BSRT} T^{(k,0)}_l = \pa T_{l+1}^{(k-1,0)} & k \geq 1\\
	Q_{BSRT} T^{(0,0)}_l=0.
	\end{cases}
\end{align}




The fundamental stress-energy tensors are local operators $T^{(k, 0)}_l$ of ghost number $k + 2l+1$.  Other tensors are obtained by descent in terms of $\dbar$-operator. These measure the response of the theory to a fluctuation of the closed-string background. 

For instance, the operator $T^{(2,0)}_0$ of ghost number $3$ can be descended to an operator $T^{(2,3)}_0$ of ghost number $0$ which measures the response of the theory to a non-commutative deformation of the Calabi-Yau geometry. If the geometry deforms by a Poisson tensor $\mu=\mu^{ij}\partial_{z_i} \wedge \partial_{z_j} $, then the Lagrangian of the open-string field theory deforms by
\begin{equation*} 
	\int T_0^{(2,3)}\wedge \bracket{\mu \vdash \Omega_X}. 
\end{equation*}

From formula \eqref{1st-order-coupling} for the coupling between open and closed string fields, we find that $T^{(k,0)}_l$ is built only from the ghost $\c$ and its $z$-derivatives, and is a sum of terms proportional to  
\begin{equation} 
	\op{Tr} \bracket{ \c^{r_1+1} (\partial  \c) \c^{r_2} (\pa c) \dots \c^{r_k} (\partial  \c)   }  
\end{equation}
where $r_1 + \dots + r_k = l$. The BRST differential relates these operators by
\begin{equation} \label{equation_brst_se} 
	Q_{\BRST} T^{(k,0)}_l= \pa T^{(k-1,0)}_{l+1}.
\end{equation}

\subsection{Large $N$ single trace operators}
A remarkable feature of holomorphic Chern-Simons theory is the following fact, which we state as a theorem:
\begin{theorem} \label{thm-HCS-operators}
	The BRST cohomology of the large-$N$ single trace operators of holomorphic Chern-Simons theory is isomorphic to the BRST cohomology of generalized stress-energy tensors $T^{(k,0)}_l$. 
\end{theorem}
This statement holds whether we work with $\mf{gl}(N)$ or $\mf{gl}(N \mid N)$. We prove it at the classical level. This theorem is  essentially equivalent to a classic result in homological algebra: the Loday-Quillen-Tsygan Theorem \cite{LQ, Ts} that relates Lie algebra cohomology at large N to cyclic cohomology. . 

We will first describe the algebra of local operators.  Let $\mc{A} \in \Omega^{0,\bullet}(\C^3 )\otimes \gl(N)$ be the master field in holomorphic Chern-Simons theory. Let $\iota_{\partial_{\zbar^i}} \A$ be the contraction with anti-holomorphic vector $\pa_{\zbar^i}$.  Given a differential form $\omega\in \Omega^{\bullet, \bullet}$, we denote by $\omega(0)$ the evaluation at $0$ of the component of $\omega$ which lies in $\Omega^{0,0}$.  A general local operator at $0$ is a polynomial in the following basic operators 
\begin{equation*} 
	(\partial_{z}^I \partial_{\zbar}^J \iota_{\partial_{\zbar}}^K \mc{A}^{i}_j )(0) 
\end{equation*}
where $I,J,K$ are multi-indices, and $i,j$ represent matrix entry in $\gl(N)$. The linearized BRST operator is $\dbar=\sum \d \zbar_i \partial_{\zbar_i}$.  If we take cohomology with respect to this linearized BRST operator, then Poincare lemma implies cancellation between $\pa_{\zbar^i}$'s and $\iota_{\partial_{\zbar^i}}$'s. We end up with cohomologies represented by the following operators
\begin{equation*} 
		\partial_{z}^I \mc{A}^{i}_j (0) =   \partial_{z}^I \c^{i}_j (0)    
\end{equation*}
which have no  $\zbar$-derivatives and $\iota_{\zbar}$ contractions. These operators are fermionic operators only involving the $z$-derivatives of the ghost field, and equipped with the usual BRST differential 
\begin{equation*} 
	Q_{\BRST} \c^{i}_j = \sum_k \c^{i}_k \c^{k}_j. 
\end{equation*}
Mathematically, this BRST cochain complex can be described as the Lie algebra cochains of $\gl(N)[[z_1,z_2,z_3]]$, where $z_1, z_2, z_3$ can be viewed as holomorphic jet coordinates at $0$.

Only $GL(N)$ invariant cochains contribute to this BRST cohomology. For $N \to \infty$, every $GL(N)$ invariant operator is a product of single-trace operators of the form
\begin{equation*} 
	\op{Tr} \bracket{ \partial_{z}^{I_1} \c \partial_{z}^{I_2} \c \dots \partial_{z}^{I_k} \c}  
\end{equation*}
where $I_1,\dots,I_k$ are multi-indices. (At finite $N$, there are trace relations which complicate the analysis). Because of the cyclic invariance of the trace, these tensors can be naturally identified as cyclically invariant linear maps $R^{\otimes k} \to \C$, where $R = \C[[z_i]]$.   

Loday-Quillen \cite{LQ} and Tsygan \cite{Ts} showed that the BRST operator on the single-trace operators is precisely the Hochschild differential on the cyclic cochain complex of $R$, with a shift in cohomological degree by $1$. We thus conclude that:
\begin{equation} 
	\text{Large $N$ single trace operators} \iso HC^\ast(R)[-1] 
\end{equation}
where $HC^\ast(R)$ is the cyclic cohomology of $R$. This is a very useful observation, which allows one to readily compute the large $N$ single trace operators of many gauge theories using techniques from homological algebra.

In the case at hand,  standard techniques of homological algebra (the Hochschild-Kostant-Rosenberg theorem and Connes' spectral sequence relating cyclic and Hochschild homology) allow us to compute $HC^\ast(R)$ very easily.  Introduce
\begin{equation} 
	\Omega^{-\bullet} = \C[[z^i, \d z^i]]	 
\end{equation}
where $\d z^i$ have cohomological degree $-1$.  This is the holomorphic de Rham complex on $\C^3$, except the functions that appear are treated as formal series and we put $k$-forms to have degree $-k$. We let $\pa$ be the holomorphic de Rham operator.  We also introduce the linear dual (continuous with respect to the adic topology):
\begin{equation*} 
	 \left( \Omega^{-\bullet}\right)^\vee.
\end{equation*}
This is spanned by $z$-derivatives of the $\delta$-function at $z^i = 0$, and its contractions with $\partial_{z^i}$. If $\iota_i$ indicates contraction with respet to $\partial_{z^i}$, then an element of $\left( \Omega^{-\bullet}\right)^\vee$ is a linear combination of
\begin{equation} 
	\partial_{z^{i_1}} \dots \partial_{z^{i_k}} \iota_{j_1} \dots \iota_{j_m} \delta_{z^i = 0} 
\end{equation}
whose value at $\alpha \in \Omega^{-\bullet}$ is $(\partial_{z^{i_1}} \dots \partial_{z^{i_k}} \iota_{j_1} \dots \iota_{j_m} \alpha)(0)$. The holomorphic de Rham differential $\pa$ on $\Omega^{-\bullet}$ induces a dual differential on $\left( \Omega^{-\bullet}\right)^\vee$ which we still denote by $\pa$. 

The Hochschild-Kostant-Rosenberg theorem implies that the cyclic cohomology group  $HC^\ast(R)$ can be computed by the cochain complex 
\begin{equation} 
	\left( \Omega^{-\bullet}\right)^\vee[[u]] \label{cochain-delta-function}
\end{equation}
with the differential $u \partial$.  We can identify elements of this complex with the components of the generalized stress-energy tensor of holomorphic Chern-Simons theory: 
\begin{equation} 
	u^l \partial_{z_{i_1}} \dots \partial_{z_{i_k}} \iota_{j_1} \dots \iota_{j_m} \delta_{z_i = 0} \leftrightarrow  \partial_{z_{i_1}} \dots \partial_{z_{i_k}} (T_l^{(m,0)})_{j_1,\dots,j_m}. 
\end{equation}
The BRST operator on the components of the generalized stress-energy tensors precisely matches the operator $u \partial$ on $ \left( \Omega^{-\bullet}\right)^\vee[[u]]$.   In this way  we see that single trace operators are, at the level of BRST cohomology, given by the components of the generalized stress-energy tensors. 

The components of the generalized stress-energy tensors arrange into multiplets $\{\partial_{z}^I T^{(k,0)}_{r-k}\}$ with  fixed $r$, which are closed under the BRST operator (see equation \eqref{stress-BRST-variation}).  We described this multiplet above in terms of the de Rham operator applied to holomorphic derivatives and contractions of $\delta_{z_i = 0}$.  It is easy to see that the de Rham cohomology of the complex \eqref{cochain-delta-function} (for $r\ge 3$) consists of simply the $\delta$-function: all derivatives and contractions cancel in cohomology.

We conclude that in the multiplet $\{\partial_{z}^I T^{(k,0)}_{r-k}\}$ for $r \ge 3$, the only operator that survives in cohomology is 
\begin{equation}
	T_r^{(0,0)} = \op{Tr} \c^{2r+1}.
\end{equation}
These are topological operators, because $\partial_{z^i}T_r^{(0,0)}$ is BRST exact. 

For $r < 3$, there are additional operators in the BRST-cohomology of this multiplet. For instance, if $r = 0$, any operator $\partial_{z^1}^{i_1}\dots \partial_{z^3}^{i_3} T_0^{(0,0)}$ is BRST closed.  For $r = 1$, the components of 
\begin{equation*}
	\pa T_0^{(1,0)}
\end{equation*}
and their derivatives are BRST closed. As we will show later, only the operators $T_r^{(0,0)}$ will further survive when we couple holomorphic Chern-Simons theory to closed string sectors. 

The above argument works in the same way when the matrix is the super Lie-algebra $\mf{gl}(N \mid N)$ and the result is identical to the $\mf{gl}(N)$ case. This is because the classical invariant theory has a natural generalization to the super Lie-algebra case \cite{Se}. We refer to \cite{CL2} for a brief discussion.

\subsection{ Operators of the coupled open-closed theory}
We have seen that, in the large $N$ limit, the space of single trace operators of the open string theory can be described in terms of closed string fields. Here we analyze what happens when we also introduce the operators of the closed string field theory. We will find that there is a remarkable cancellation which implies the coupled theory is purely topological and has no degrees of freedom.      (This effect is only visible at infinite $N$. At finite $N$ it is spoiled by trace relations).

Recall that the $B$-model closed string states are given as the $S^1$-equivariant cohomology of the Hilbert space of the $B$-model TFT.  The parameter $u$ in the closed string states is the equivariant parameter.   We can also consider the \emph{localized} $S^1$-equivariant cohomology of the $B$-model TFT.  This is obtained by inverting $u$, giving us $\PV^{\bullet,\bullet}((u))$.

We could try to build a (rather trivial) string-field theory based on these localized string states.  Without this localization, the string-field theory is holomorphic on the space-time $X$.  After we localize, it is topological.  This is because the action of divergence-free holomorphic vector fields is BRST exact, due to the BV relation \eqref{eqn-BV}:
\begin{equation} 
	\mc{L}_V\alpha =[V, \alpha]= Q ( u^{-1}V \wedge \alpha)- u^{-1}V\wedge Q(\alpha), \quad Q=\dbar +u\pa_\Omega
\end{equation}
for any $\alpha \in \PV^{\bullet,\bullet}((u))$ and any divergence-free holomorphic vector field $V$ (i.e. $QV=0$). In other words, the following Carten homotopy formula holds
\begin{equation}
\quad \mc{L}_V=[Q, u^{-1}V]. 
\end{equation}

At the level of BRST cohomology, the local operators of this string-field theory are generated by an infinite sequence of topological operators $T_l$, where $T_l$ measures the field in $u^{-l-1} \PV^{3,0}$. This is due to the simple fact that locally on the manifold $X$ 
\begin{equation}
 H (\PV^{\bullet, \bullet}((u)), \dbar+u\pa_\Omega)\iso \C((u))[ \Omega^{-1} ]. 
\end{equation}
Here $\Omega^{-1}=\pa_{z^1}\wedge \pa_{z^2}\wedge \pa_{z^3}\in \PV^{3,0}$ is the holomorphic polyvector whose contraction with $\Omega$ is $1$. The operator $T_l$ is of ghost number $2l+1$.  

The cancellation between the open and closed operators at large $N$ is the following:
\begin{theorem} \label{thm-BRST-cohomology}
	At the level of BRST cohomology, the operators of the coupled open-closed theory at large $N$ are isomorphic to the operators of the localized closed-string field theory. That is, the algebra is generated by topological operators $T_l$ of ghost number $2l+1$, for $l \in \Z$. 
\end{theorem}

The proof of this result is purely algebraic. Recall that we have operators $T^{(k,m)}_{l}$ ($l\geq 0$) of stress-energy tensors from open string sector  described in Section \ref{sec-stress-tensors}. Now we introduce operators $T^{(k,m)}_{l}$ ($l<0$) from closed string sector as follows. $T^{(k,m)}_{l} \in \Omega^{k,m}$ depends only on closed string fields in $u^{-l-1} \PV^{3-k,m}$ and is given by 
\begin{equation}
T^{(k,m)}_{l} = \mu\vdash \Omega, \quad\text{for}\quad u^{-l-1} \mu \in u^{-l-1} \PV^{3-k,m}, l<0. 
\end{equation}
These operators capture the same information as the closed string fields. The linearized BRST operator $Q=\dbar+u\pa_\Omega$ on closed string fields induces a linearized BRST operator on closed string operators $T^{(k,m)}_{l}$ by 
\begin{align}\label{stress-BRST-variation-KS}
  \begin{cases}  Q T_l^{(k,m)} = \dbar T_l^{(k,m-1)}+\pa T_{l+1}^{(k-1,m)} & k,m\geq 1, l<-1\\
             Q T_{-1}^{(k,m)} = \dbar T_{-1}^{(k,m-1)} & k,m\geq 1\\
        Q T^{(0,m)}_l =\dbar T_l^{(0,m-1)} & m\geq 1, l<0\\
	Q T^{(k,0)}_l = \pa T_{l+1}^{(k-1,0)} & k \geq 1, l<-1\\
	Q T^{(k,0)}_{-1} = 0 & k \geq 1\\
	Q T^{(0,0)}_l=0 & l<0
	\end{cases}
\end{align}
We observe that they are similar to the BRST transformation \eqref{stress-BRST-variation} of stress-energy tensors from open string sector, except for a discrepancy at $T_{-1}^{(\bullet, \bullet)}$. This is precisely corrected by the coupling of open-closed sectors!

To see this, let $I^{(1)}$ denote the first order coupling of holomorphic Chern-Simons fields with Kodaira-Spencer gravity described in Section \ref{section-coupliing}. There are extra terms in the BRST transformation of closed string operators  by 
\begin{equation}
T^{(k,m)}_{l}\to \{ I^{(1)}, T^{(k,m)}_{l}\}_c. 
\end{equation}
Here $\{-,-\}_c$ is the BV anti-bracket  for Kodaira-Spencer fields. By \eqref{KS-BV-bracket}, this is only nontrivial for $T^{(k,m)}_{-1}$. Using formula \eqref{1st-order-coupling}, we find 
\begin{equation}\label{BRST-connecting}
\{ I^{(1)}, T^{(k,m)}_{-1}\}_c= \begin{cases}
\pa T_{0}^{(k-1,m)}  & k>0\\
0  & k=0. 
\end{cases}
\end{equation}
where $T_{0}^{(k-1,m)}$ is the stress-energy tensor from open string sector. 

This is the BRST transformation connecting  open and closed string sectors! 

Combining \eqref{stress-BRST-variation} \eqref{stress-BRST-variation-KS} \eqref{BRST-connecting}, we find the following BRST transformations for operators in the coupled open-closed theory
\begin{equation}
Q_{\BRST} T_l^{(k,m)} =  \dbar T_l^{(k,m-1)}+\pa T_{l+1}^{(k-1,m)}. \label{BRST-variation-coupled}
\end{equation}
Now $l\in \Z$ runs over all integers, and it is understood that $T_l^{(k,m)}=0$ if $k<0$ or $m<0$. 

Using Theorem \ref{thm-HCS-operators}, we find that at the level of BRST cohomology, local operators in the coupled open-closed theory are generated by components of $\{T_l^{(k,m)}\}_{l\in \Z}$ and their derivatives. Formula \eqref{BRST-variation-coupled} implies that all derivatives and form factors will cancel in BRST cohomology. Therefore the BRST cohomology will then consist only of the operators $T_{l}^{(0,0)}$ of ghost number $2l+1$, for $l \in \Z$.  The derivatives of these operators are zero in cohomology, so that these are topological operators at the level of BRST cohomology\footnote{One might worry that there are further corrections to the BRST operator which take some $T$ to a product of $T$'s.  One can show that this is not possible, for dimensional reasons. The point is that any such term in the BRST operator must occur for a world-sheet of Euler characteristic zero, otherwise it would be dimensionful which is impossible since all operators are dimensionless.  It is not hard to show directly that world-sheets of Euler characteristic zero don't contribute any further terms.}. 

We conclude that at the level of BRST cohomology,  the algebra of operators of the coupled open-closed theory is simply the polynomial algebra generated by the topological operators  $\{T_{l}^{(0,0)}\}_{l\in \Z}$.  This proves Theorem \ref{thm-BRST-cohomology} with the identification $T_l=T_{l}^{(0,0)}$.

\subsection{A comment on counting loops}

We have explained how the Green-Schwarz mechanism leads to cancellation of one-loop anomalies.  Now we will explain one of the main points of this note: \emph{anomalies at higher loops are also cancelled, and counter-terms at all loops are fixed uniquely.}

Let us first comment on how we are counting loop number.  As in any open-closed string field theory, we can associate the topological type of a Riemann surface with boundary to any Feynman diagram.  Closed-string vertices are viewed as pairs of pants, and open-string vertices are viewed as discs with marked points on the boundary.  Let $\gamma$ be such a diagram, and $\Sigma_{\gamma}$ be the corresponding genus $g$ Riemann surface with $h$ boundary components and $n$ closed-string marked points.  The loop number of $\gamma$ is
\begin{equation} 
	2g-2 + h + n  +1  = - \chi(\Sigma_{\gamma}) + n +1 
\end{equation}
where $\chi(\Sigma_{\gamma})$ is the topological Euler characteristic.  In other words, we count a closed string marked point as the same loop number as an open string boundary component. Such a graph comes with a factor of $\lambda^{2g-2+h+n}=\lambda^{- \chi(\Sigma_{\gamma}) + n}$ where $\lambda$ is the string coupling constant. Note that $\chi(\Sigma_\gamma)-n$ is the Euler characteristic of the surface obtained from $\Sigma_\gamma$ with closed string marked point deleted. 

In this counting, the open-string interaction occurs with a coefficient of $\lambda^{-1}$, because a disc has Euler characteristic $1$. The open string propagator occurs with coefficient $\lambda$, because when we glue to surfaces along an interval in their boundaries the Euler chacteristic decreases by one. The closed-string propagator has coefficient $\lambda^{0}$, as gluing surfaces along a common boundary preserves Euler characteristic.  The closed-string vertex has coefficient $\lambda$, as the sphere with three punctures has Euler characteristic $-1$. 

The tree-level graphs are given by disks without closed string insertions, i.e., $g=0, h=1, n=0$. They are computed by tree-level Feymann diagrams in holomorphic Chern-Simons theory. It satisfies the BV quantum master equation \cite{Z2} at the level of $g=0, h=1, n=0$ as a consequence of the classical master equation \eqref{CME-HCS}. 

The one-loop graphs are those with a power of $\lambda^0$. There are two cases with non-trivial insertions: $g=0, h=1, n=1$ and $g=0, h=2, n=0$. The first case is computed by tree-level Feynman diagrams with one vertex by the first order coupling $I^{(1)}$ as in Section \ref{section-coupliing} and other vertices by the holomorphic Chern-Simons interaction. It satisfies the BV quantum master equation \cite{Z2} at the level of $g=0, h=1, n=1$ by the construction of $I^{(1)}$ (more precisely by the linearized BRST transformation property: $QI^{(1)}+\delta_{HCS}I^{(1)}=0$). 

The Riemann surface associated to the second case $g=0, h=2, n=0$  is a cylinder. The BV quantum master equation \cite{Z2} at this level is pictured as the degeneration of the cylinder
\[
\begin{tikzpicture}[scale=.15]
\draw[thick](0,0)ellipse(1 and 3) (0,3)to(10,3)(0,-3)to(10,-3);
\draw[thick](10,3)arc(90:-90:1 and 3)(10,-3);
\draw[dashed,thick](10,3)arc(90:270:1 and 3)(10,-3);
\draw(0,0)node[left]{\Huge{$\partial($}} (11,0)node[right]{\Huge{$)=$}}(0,-5)node{};
\end{tikzpicture}
\begin{tikzpicture}[scale=.15]
\draw[thick](0,0)ellipse(1 and 3) (0,3)to(10,3)(0,-3)to(10,-3)
    (-4,0)ellipse(3 and 3);
\draw[thick](10,3)arc(90:-90:1 and 3)(10,-3);
\draw[dashed,thick](10,3)arc(90:270:1 and 3)(10,-3);
\draw(11,0)node[right]{\Huge{$+$}}(0,-5)node{};
\end{tikzpicture}
\begin{tikzpicture}[scale=.15]
\draw[thick](0,0)ellipse(3 and 5);
\draw[thick,fill=white](-3,0)ellipse(3 and 3);
\draw(3,0)node[right]{\Huge{$+$}};
\end{tikzpicture}
\begin{tikzpicture}[scale=.15]
\draw[thick](0,0)ellipse(1 and 3) (0,3)to(10,-3)(0,-3)to(10,3);
\draw[thick](10,3)arc(90:-90:1 and 3)(10,-3);
\draw[dashed,thick](10,3)arc(90:270:1 and 3)(10,-3);
\draw(11,0)node[right]{\Huge{$$}}(0,-5)node{};
\end{tikzpicture}
\]
These are exactly the graphs that appeared in our discussion of the Green-Schwarz mechanism. 

If we write $I_{g,h,n}$ for the effective interaction of genus $g$ with $h$ boundary components and $n$ closed-string marked points, then the above picture reads as the following master equation
\begin{equation}
{QI_{0,2,0}+\fbracket{I_{0,1,0},I_{0,2,0}}_{o}+\Delta_{o} I_{0,1,0}+ {1\over 2}\fbracket{I_{0,1,1},I_{0,1,1}}_{c}=0}.
\end{equation}
Here $\{-,-\}_o$ is the BV bracket in the open string sector (HCS), $\{-,-\}_c$ is the BV bracket in the closed string sector (KS gravity), and $\Delta_o$ is the BV operator in the open string sector. This master equation has an anomaly to solve if the gauge Lie algebra is $gl(N)$, and anomaly free if the gauge Lie algebra is $gl(N|N)$, as discussed in section \ref{sec-GS}.

This counting of loop number has the following nice feature. If $\mu$ is a polyvector field on $\C^d$, let us give it a dimension based on the dimension of the differential form $\mu \vdash \Omega$.  Then, a field $\mu_{\bar j_1\cdots \bar j_l}^{i_1\cdots i_k}$ in a component of $\PV^{k,l}$ has dimension $d-k+l$, and charge $d-k-l$ under the diagonal $U(1)$ in $U(d)$.  Similarly,  for a HCS field $\A_{\bar i_1\cdots \bar i_k}$ in a component of $\Omega^{0,k}\otimes \g$, we give it dimension $k$ and $U(1)$ charge $-k$. If we give the string coupling constant $\lambda$ dimension $d$ and $U(1)$ charge $d$, then the open-closed Lagrangian is dimensionless and $U(1)$ invariant.   

For example, the interaction of holomorphic Chern-Simons in complex dimension $3$ is a linear combination of expressions of the form 
$$
    \int d^3z d^3 \bar z \A_{\bar I_1}\A_{\bar I_2}\A_{\bar I_3} , \quad \text{with}\quad |I_1|+|I_2|+|I_3|=3. 
$$
It has dimension $6-|I_1|-|I_2|-|I_3|=3$. This is because the volume density $d^3z d^3\bar z$ has dimension $6$, and the minus sign is because a functional of a field has the opposite dimension to the corresponding field. If we take into count the string coupling constant $\lambda$, then $\lambda^{-1}\int  \A_{\bar i_1}\A_{\bar i_2}\A_{\bar i_3} d^3z d^3 \bar z$ is dimensionless. 

For another example,  let $\mu\in \PV^{k,l}$ and consider its first order coupling $I^{(1)}_{\mu}[\A]$ \eqref{1st-order-coupling} with HCS fields. It is given by a linear combination of the form
$$
\int d^3z d^3 \bar z \mu_{\bar j_1\cdots \bar j_l}^{i_1\cdots i_k} \A_{\bar I_{0}}\pa_{z^{i_1}}\A_{\bar I_1}\cdots \pa_{z^{i_k}}\A_{\bar I_k}, \quad \text{with}\quad |I_0|+\cdots+|I_k|=3-l.
$$
Since each derivative $\pa_{z^i}$ has dimension $-1$, this interaction has dimension $(k-3-l) - (|I_0|+\cdots+|I_k|+k)+6=0$ and it comes with a coupling constant $\lambda^0$ as expected. 

\subsection{Quantization and anomaly cancellations at higher loops}\label{sec-quantization}
Now let us turn to the analysis of higher-loop graphs. We have seen that coupling the open and closed $B$-model yields a theory that, at large $N$, is purely topological.  In particular, all local operators are of dimension $0$ (for this to be true, it is important that we count the dimension of polyvector fields as described above). This leads to a remarkable cancellation of potential higher-loop anomalies:  beyond the leading order in the expansion in the string coupling,  the coupled theory has \emph{no possible counter terms or gauge anomalies}. 

The reason is the following.  As we have seen,  the string coupling constant is dimensionful, so that possible counter-terms or gauge anomalies beyond the one loop level must be dimensionful.  Theorem \ref{thm-BRST-cohomology} implies that there are, however, no dimensionful counter-terms or potential gauge anomalies in the BRST cohomology. Any counter-terms or gauge anomalies arise by descent from local operators $T_l$, which are all dimensionless.

As a consequence, it shows that the dynamics of Kodaira-Spencer gravity are fully recovered from the large N holomorphic Chern-Simons theory! For example, consider the following BV master equation \cite{Z1} at genus $2$ in the closed string sector

\[
\begin{tikzpicture}[scale=.14]
\draw[thick]plot [smooth,tension=1] coordinates
    {(1,0)(3.5,3)(10,3)(17,3)(17,-1)(10,-2)(3,-3)(1,0)};;
\draw[thick, fill=white](4,0)to[bend right](7,0);
\draw[thick, fill=white](4.5,0)to[bend left](6.5,0);

\draw[thick, fill=white](4,0)to[bend right](7,0);
\draw[thick, fill=white](4.5,0)to[bend left](6.5,0);
\draw[thick, fill=white](4+6,0+1)to[bend right](7+6,0+1);
\draw[thick, fill=white](4.5+6,0+1)to[bend left](6.5+6,0+1);

\draw(1,0)node[left]{\huge{$\partial($}} (18,0)node[right]{\Huge{$)=$}};
\draw (16,1)node{\tiny{$\times$}}(12,-1)node{\tiny{$\times$}}(6,1.5)node{\tiny{$\times$}};
\end{tikzpicture}
\begin{tikzpicture}[scale=.15]
\draw[thick]plot [smooth,tension=1] coordinates
    {(1,0)(3.5,3)(10,3)(17,3)(17,-1)(10,-2)(3,-3)(1,0)};;
\draw[thick, fill=white](4+6,0+1)to[bend right](7+6,0+1);
\draw[thick, fill=white](4.5+6,0+1)to[bend left](6.5+6,0+1);
\draw(18,0)node[right]{\Huge{$+$}};
\draw (16,1)node{\tiny{$\times$}}(12,-1)node{\tiny{$\times$}}(6,1.5)node{\tiny{$\times$}};

\draw[thick] plot [smooth,tension=1] coordinates{(1,0) (5,1) (5,-1) (1,0)};
\end{tikzpicture}
\begin{tikzpicture}[scale=.15]
\draw[thick](5.5,0)ellipse(4.5 and 3);
\draw[thick](14,0)ellipse(4 and 3);
\draw[thick, fill=white](4,0)to[bend right](7,0);
\draw[thick, fill=white](4.5,0)to[bend left](6.5,0);

\draw[thick, fill=white](4,0)to[bend right](7,0);
\draw[thick, fill=white](4.5,0)to[bend left](6.5,0);
\draw[thick, fill=white](4+8,0+1)to[bend right](7+8,0+1);
\draw[thick, fill=white](4.5+8,0+1)to[bend left](6.5+8,0+1);
\draw(18,0)node[right]{\Huge{$+$}};
\draw (16,1)node{\tiny{$\times$}}(12,-1)node{\tiny{$\times$}}(6,1.5)node{\tiny{$\times$}};
\end{tikzpicture}
\begin{tikzpicture}[scale=.15]
\draw[thick]plot [smooth,tension=1] coordinates
    {(1,0)(3.5,3)(10,3)(15,3)(15,-1)(10,-2)(3,-3)(1,0)};;
\draw[thick, fill=white](4,0)to[bend right](7,0);
\draw[thick, fill=white](4.5,0)to[bend left](6.5,0);

\draw[thick, fill=white](4,0)to[bend right](7,0);
\draw[thick, fill=white](4.5,0)to[bend left](6.5,0);
\draw[thick, fill=white](4+6,0+1)to[bend right](7+6,0+1);
\draw[thick, fill=white](4.5+6,0+1)to[bend left](6.5+6,0+1);
\draw (12,-1)node{\tiny{$\times$}}  (-3,2)node{\tiny{$\times$}}(-3,-2)node{\tiny{$\times$}};
\draw[thick](-2,0)ellipse(3 and 3) (-5,0)to[bend left=-60](1,0);
\draw[dashed,thick](-5,0)to[bend left=60](1,0);
\end{tikzpicture}
\]
It reads as the following equation (using the notation $I_{g,h,n}$ as above)
\begin{equation}
  QI_{2,0,n}+\Delta_{c} I_{2,0,n+2}+{1\over 2}\sum_{n_1+n_2=n}\{I_{1,0,n_1+1}, I_{1,0,n_2+1}\}_c+\sum_{n_1+n_2=n}\{I_{0,0,n_1+1}, I_{0,0,n_2+1}\}_c=0
\end{equation}
Here $\Delta_c$ is the BV operator in the closed string sector. Such equation is solved without ambiguity (up to gauge equivalence) as part of the recursive process of constructing higher loop graphs via the BRST cohomology analysis. 

It seems like this argument might only work for $N = \infty$, because the cancellation between open and closed string operators only holds when we do not impose trace relations on the open-string side. However, the argument can be applied for all values of $N$; see \cite{CL2} for details.   To do this, we should consider gauge theories defined for all $\mf{gl}(N)$ (or $\mf{gl}(N|N)$) in a uniform way.  The Feynman diagrams of such gauge theories are ribbon-graphs (also known as fat graphs).  The amplitude of a ribbon graph can be evaluated for a $\mf{gl}(N)$ (or $\mf{gl}(N|N)$) gauge theory for any $N$.   In this set-up,  local operators, counter-terms, etc. are all given as products of cyclic words of the gauge fields, without imposing trace relations.  The cancellation between open and closed string theory operators holds in this setting since we have not imposed trace relations.  This implies  the higher-loop anomaly cancellations  for all  $N$. In particular, if we work with $\mf{gl}(N|N)$ where one-loop anomaly cancellation holds by Green-Schwarz mechanism, then we have established quantization of open-closed B-model at all loops in perturbation theory.

\section{The type I topological string}

\subsection{Holomorphic Chern-Simons for orthogonal groups} \label{sec-HCS-SO}
So far, we have discussed holomorphic Chern-Simons theory for the Lie algebras $gl(N)$ or $\gl({N\mid N})$.  In this section we will explain a parallel analysis in the case of $\mf{so}(N)$. 

Holomorphic Chern-Simons theory for Lie algebras other than $\gl(N)$ can not be coupled to the full Kodaira-Spencer theory.  The Poisson tensor field of Kodaira-Spencer theory makes the space-time non-commutative, and one can not have a $\g$-bundle on a non-commutative space-time unless $\g$ is of type $A$. For example,   if $\mc{A} \in \Omega^{0,\ast}(\C^n)\otimes \g[1]$ denotes the holomorphic Chern-Simons super-field, then the coupling to a Poisson tensor $\pi^{ij}$ takes the form
\begin{equation*} 
	\int \phi_{abc} \mc{A}^a \partial_{z^i} \mc{A}^b \partial_{z^j} \mc{A}^c \pi^{ij} \Omega 
\end{equation*}
where $a,b,c$ are Lie algebra indices and $\phi_{abc}$ is some invariant cubic tensor.  For this to be non-zero, $\phi_{abc}$ must be symmetric in the $b$ and $c$ indices.  There is only such a tensor when $\mf{g}$ is of type A.

There is, however, a variant of Kodaira-Spencer theory which can be coupled to holomorphic Chern-Simons theory for general Lie algebras. In \cite{CL2} we called this $(1,0)$ Kodaira-Spencer theory, because of its relationship to the $(1,0)$ tensor multiplet in $6$ dimensions.  A better name for this system could be the type I topological string, because its relationship to the ordinary $B$-model is similar to the relationship of physical type I string theory to the IIB string theory. 

The fields of Kodaira-Spencer theory are the $SO(2)$-equivariant states of the $B$-model TFT.  We took $SO(2)$-equivariant cohomology because we were interested in operators that, by descent, correspond to Lagrangians that can be defined on any oriented Riemann surface.

For type I Kodaira-Spencer theory, we would like to consider those operators that descend to Lagrangians defined on a surface without a choice of orientation. Any local operator can be descended twice to give a $2$-form valued local operator.  On a Riemann surface without an orientation, we can not integrate a two-form. Rather, we can integrate a two-form twisted by the orientation local system, which is a flat rank $1$ vector bundle whose monodromy is given by the change of orientation. 

This tells us that in the type I theory, we should not just consider $O(2)$-equivariant local operators: these will descend to give $2$-form valued operators. Instead, we should consider $O(2)$-equivariant local operators after we have twisted by the determinant representation of $O(2)$.

The $O(2)$-equivariant cohomology is simply the $\Z/2$-fixed points of the $SO(2)$-equivariant cohomology.   The $O(2)$-equivariant cohomology after twisting by the determinant representation is the subspace of the $SO(2)$-invariant cohomology on which $\Z/2$ acts by $-1$.

One can show that $\Z/2$ action has eigenvalue $1$ on the space $u^l \PV^{k,\bullet}$ for $k+l$ even, and eigenvalue $-1$ on these spaces when $k+l$ is odd.  Thus, the $O(2)$-equivariant cohomology, twisted by the determinant representation of $O(2)$, is given by
\begin{equation}
	\sum_{k+l \text{ odd}, l\geq 0} u^l \PV^{k,\bullet}.
\end{equation}
The BRST differential is, as before, $\dbar +u\partial_\Omega$.  These are the fields of type I Kodaira-Spencer theory.

At the level of open-string, for a field $\mc{A} \in \Omega^{0,\ast}(X)\otimes \mf{gl}_N[1]$, the involution coming from an orientation-reversing symmetry of the world-sheet sends $\mc{A} \mapsto \mc{A}^T$.  As before, the fields of the open-string field theory must be quantities that we can integrate over the boundary of a non-orienatble surface, and so must transform by $-1$ under the $\Z/2$ action.  Therefore the the open-string field theory is holomorphic Chern-Simons theory for $\mf{so}_N$. 

The closed-string field theory couples to holomorphic Chern-Simons theory for $\mf{so}(N)$ for any $N$.  The formula for the coupling is the same as that described in Section \ref{section-coupliing}, which makes sense for the gauge Lie algebras $\mf{so}(N)$.  In fact, the  coupling with the field
\begin{equation*}
	\mu \in u^l \PV^{k,\bullet}
\end{equation*}
involves a sum of traces of $2l+k+1$ copies of the master field $\mc{A}$ and its derivatives.  One can show that this expression changes by $(-1)^{k+l+1}$ when we apply the symmetry that sends $\mc{A}$ to $-\mc{A}^T$ while reversing the order of the trace, so that only the fields in $u^l \PV^{k,\bullet}$ with $k+l$ odd couple non-trivially.  For instance, the field in $u^l \PV^{0,\bullet}$ couples to $\op{Tr} (\A^{2l+1})$, which is zero for matrices in $\mf{so}(N)$  when $l$ is even.

The analysis of the large $N$ single trace operators for $\mf{so}(N)$ proceeds in a similar way to that for $\mf{gl}(N)$.  The only difference is that single trace operators for $\mf{so}(N)$ are represented by the $\Z/2$ anti-invariants of the cyclic cohomology group that describes the $\mf{gl}(N)$ operators.  The $\Z/2$ action is the same as that on the fields of Kodaira-Spencer theory, so we conclude that in the large $N$ limit, all operators in BRST cohomology are given by the generalized stress-energy tensors that are responses to type I Kodaira-Spencer fields. This is similar to the discussion in Section \ref{sec-stress-tensors}.

In the case of $\mf{so}(N)$, because only half of the fields of Kodaira-Spencer theory contribute, the description of the operators is a little simpler.  After passing to BRST cohomology, the list of operators consists of 
\begin{equation}
 T^{(0,0)}_{2l-1} = \op{Tr} \c^{4l-1}
\end{equation}
for $l \ge 1$, and the holomorphic stress-energy tensor
\begin{equation}
	T^{(1,0)}_0 = \op{Tr}\c \partial\c.
\end{equation}
The operators $T_{2l-1}$ for $l \ge 2$ are, at the level of BRST cohomology, topological operators, whose derivatives are BRST exact.  The stress-energy tensor $T^{(1,0)}_0 $ is related to $T_1^{(0,0)}$ as before, by
\begin{equation*}
	Q_{\BRST} T^{(1,0)}_0= \partial T_1^{(0,0)}.
\end{equation*}
Thus the space of single-trace operators in this case consists of the stress-energy tensor for a holomorphic theory defined on Calabi-Yau manifolds, together with an infinite tower of topological operators.

\subsection{Anomaly cancellation in type I topological string in $3$ complex dimensions}
We will show that the Green-Schwarz mechanism cancels the anomalies for the type I topological string in $3$ complex dimensions when coupled to holomorphic Chern-Simons for the group $SO(8)$.  This anomaly cancellation extends to all orders in the loop expansion.

As in the ordinary $B$-model, the open-string anomaly takes the same form     
\begin{equation}
\int \Tr_{\ad} \A (F_{\A})^3.
\end{equation}
where we take the trace in the adjoint representation of $\mf{so}(N)$.  

The closed-string contribution to the one-loop anomaly is given by
\begin{equation} 
	\int \Tr_{fun}(\A F_{\A}) \Tr_{fun}(F_{\A}^2 ).  
\end{equation}
This is the only term that appears when we use the type I topological string: it is the one associated to closed-string fields in $\PV^{1,\ast}$.  Here by $\Tr_{fun}$ we simply mean the Killing form on our Lie algebra $\g$, normalized according to the trace in some fundamental representation.

The Green-Schwarz anomaly can be cancelled only if
\begin{equation} 
	\op{Tr}_{ad} (X^4) \propto \left( \op{Tr}_{fun}(X^2)\right) ^2 
\end{equation}
where $X \in \g$.  

For the classical Lie algebras, the space of quartic invariant polynomials is always two dimensional, with the following exceptions. For $\g = \mf{sl}_2$, $\g = \mf{sl}_3$, the space of quartic invariant polynomials is one dimensional.  For $\g = \mf{so}_8$, the space of quartic invariant polynomials is $3$ dimensionals.  

For the exceptional groups, the space of quartic invariant polynomials is always one dimensional. 

We find that the one-loop Green-Schwarz anomaly can be cancelled for all exceptional groups, $\mf{sl}_2$, $\mf{sl}_3$. For the other classical groups, the anomaly can be cancelled only in those situations when $\op{Tr}_{ad} X^4$ is not an independent generator of the ring of invariant polynomials.

It turns out that this happens only for $\mf{g} = \mf{so}_8$.  We have the following identity for $\g = \mf{so}_n$:
$$
\Tr_{ad} (X^4)=(n-8)\Tr_{fun}(X^4)+3 (\Tr_{fun}(X^2))^2
$$
This means that the one-loop anomaly vanishes if $n = 8$.

More generally, if we take for $\mf{g}$ the superalgebra $\mf{osp}(n\mid m)$, the same identity holds with $n-m$ playing the role of $n$.  We find the anomaly cancels for $\mf{g} = \mf{osp}(n+8 \mid n)$.

For type $A$, it is easy to verify that the anomaly does not cancel for $\mf{g} = \mf{sl}_n$ for any $n > 3$.

\subsection{Anomaly cancellation beyond one loop}
Let us now briefly explain how higher loop anomalies are cancelled. We will take the holomorphic Chern-Simons gauge group to be $\mf{osp}(N + 8 \mid N)$.  Sending $N \to \infty$ has the effect of removing the trace relations from the open-string algebra. Following our analysis earlier, if we couple the type I topological string to holomorphic Chern-Simons theory for $\mf{osp}(N + 8 \mid N)$ for $N \to \infty$, we find that only topological operators survive on the space-time.  at $N \to \infty$ 

Essentially the same argument shows that, for $N$ large, the theory coupling $\mf{so}(N)$ holomorphic Chern-Simons with the type I closed string field theory is topological. The topological operators are half of those  present in the ordinary $B$-model: namely, the operators $T_{2l-1}$, for $l \in Z$, of ghost number $4l-1$. These are exactly the operators we get when we perform localized equivariant cohomology of the $B$-model TFT with respect to the group $O(2)$ instead of $SO(2)$.  

Similar argument as in Section \ref{sec-quantization} for $\gl({N|N})$ case shows that anomalies cancel beyond the one-loop setting as well. As a result,  we have established quantization of our twisted type I topological string at all loops in perturbation theory.

\subsection{Anomaly cancellation in the type I topological string in $10$ dimensions and $SO(32)$}

The original Green-Schwarz anomaly cancellation occurs for the $10$-dimensional type I superstring.  We argued in \cite{CL3} that a supersymmetric localization of the space-time theory of physical type IIB string theory can be expressed in terms of the $B$-model on Calabi-Yau $5$-folds.   In particular, we argued that a twist of type IIB supergravity on $\R^{10}$ is given by the closed-string field theory of the topological $B$-model  on $\C^5$.  In the open string sector, Baulieu \cite{Bau} showed that a supersymmetric localization of the theory on a $D9$ brane is equivalent to holomorphic Chern-Simons theory on $\C^5$.

In  complex dimension $5$, there is a ghost-number anomaly for the topological string. This means that we always need to insert fields of non-zero ghost number to get non-trivial correlators.  Equivalently, we should give the string coupling constant $\lambda$ ghost number $-2$ to put the action to have ghost number zero.  In this dimension, we should always treat our fields in the BV formalism, so the fundamental field of holomorphic Chern-Simons theory is the master field $\mc{A} \in \Omega^{0,\bullet}(\C^5) \otimes \g[1]$.      

In \cite{CL2}, we showed that there is a unique quantization of coupled Kodaira-Spencer theory and $\mf{gl}(N \mid N)$ holomorphic Chern-Simons theory, for all $N$, in  complex dimension $5$ as well. The proof is identical to the one sketched above in the case of complex dimension $3$. 

Here we will generalize this analysis to type I topological strings, and show how the Green-Schwarz mechanism considered here is related to that in physical string theory.

The space-time fields associated to Type I string theory are obtained from those of type IIB by looking at the fixed points of a $\Z/2$ action on the space of fields.  This procedure is compatible with supersymmetry, and so can be applied to compute the supersymmetric localization of the space-time theory of type I string theory. The closed-string fields of the localized type IIB twisted theory are $\PV^{\bullet,\bullet}(\C^5)[[u]]$. We propose, as before, that $\Z/2$ acts by $(-1)^{k+l+1}$ on the fields in $\PV^{k,\ast}(\C^5)u^l$.  Therefore our proposal for the fields of twisted type I supergravity are
\begin{equation} 
	\sum_{k+l \text{ odd}, l\geq 0 }\PV^{k,\ast}(\C^5) u^l. 
\end{equation}
The fundamental field is, as before, the Beltrami differential $\beta \in \PV^{1,\ast}(\C^5)$.  We also have a dynamical field in $\PV^{3,\ast}(\C^5)$.  The fields in $\PV^{k,\ast}(\C^5)u^l$, where $k+l \ge 5$, provide only consistent topological degrees of freedom and are not so important.  

If we have a stack of $N$ $D9$ branes, the closed-string fields in type I should be obtained by applying an involution corresponding to reversing the order of the open strings. As in Section \ref{sec-HCS-SO},  this involution is simply $A \to A^T$. The open-string fields in type I consist of the fields that are odd under this involution, so they give  holomorphic Chern-Simons theory with gauge Lie algebra $\mf{so}(N)$. More generally, if we start with $N$ $D9$ and $M$ anti-$D9$, we will get holomorphic Chern-Simons theory for  $OSp(N \mid M)$.

Next we will show that the open-string anomaly can be cancelled only if we use $OSp(32+N \mid N)$.  Our result is in some ways stronger than the original result of Green-Schwarz, because it holds to all orders in perturbation theory and not just at leading order. It would be very interesting to see whether our higher-loop anomaly cancellation can be applied to physical theories. 

In complex dimension $5$, the one-loop anomaly for holomorphic Chern-Simons theory with gauge Lie algebra $\mf{g}$ is 
\begin{equation} 
	\int \op{Tr}_{ad}( \A F(\A) ^5 ) = \int	\op{Tr}_{ad}( \A (\partial \A)^5 )  
\end{equation}
where the trace is taken in the adjoint representation. This is similar to complex dimension $3$ case and was computed in \cite{CL2} for arbitrary dimensions in the current context. 

If we couple our twisted type I super-gravity minimally to holomorphic Chern-Simons theory with gauge Lie algebra $\g$, then only the fields $\beta \in \PV^{1,\ast}(\C^5)$ and $\eta \in \PV^{0,\ast}(\C^5) u$ are coupled.  The field $\beta$ deforms the complex structure, and is coupled by
\begin{equation*} 
	\int  (\beta\vdash \op{tr} \mc{A} \pa \mc{A})\wedge \Omega.
\end{equation*}
Here $\op{tr}$ indicates the trace in the vector representation and $\Omega=d^5 z$ is the holomorphic volume form. The field $\eta$ deforms the volume form, and is coupled by 
$$
\int (\eta \op{tr} \mc{A}^3)\wedge \Omega.
$$

These are similar to the discussion in Section \ref{section-coupliing}. Let $I^{(1)}$ collect such first order couplings.

The closed-string BV anti-bracket $\{-,-\}_c$ in complex dimension $5$ is similar to equation \eqref{KS-BV-bracket}, and the only nontrivial BV anti-bracket is between fields in $\PV^{k, \bullet}$ and $\PV^{4-k, \bullet}$. It encodes the integral kernel of the divergence operator $\pa_\Omega$. If we only turn on fields $\beta, \eta$ as above, then $\{I^{(1)}, I^{(1)}\}_{c}$ vanishes, because the closed-string fields $\beta, \eta$  have trivial BV bracket with each other.  To cancel the open-string anomaly, we need to add a term where the fields in $\PV^{3,\ast}(\C^5)$ are coupled since the BV anti-bracket of a field in $\PV^{3,\ast}$ with one in $\PV^{1,\ast}$ is non-trivial.

There is a potential coupling whereby $\mu\in \PV^{3,\ast}$ is coupled by 
\begin{equation*} 
	\int \left(\mu\vdash\op{tr} \left( \mc{A} \partial \mc{A} \partial  \mc{A} \partial \mc{A}\right)\right)\wedge \Omega.  
\end{equation*}
(To make this closed under the linearized BRST operator, we also need to couple the fields in $u^l \PV^{3-l, \bullet}$ to $l+4$ many $\mc{A}$'s, as explained in Section \ref{section-coupliing} and \ref{sec-stress-tensors}). If we incorporate this coupling into $I^{(1)}$ , then 
\begin{equation} 
	\{I^{(1)}, I^{(1)}\}_c \propto \int \op{tr} ( \A \partial \A) \op{tr} ( (\partial \A)^4). 
\end{equation}
We find that the open and closed-string anomalies can be cancelled only if 
\begin{equation} 
	\int \op{tr} ( \A \partial \A) \op{tr} ( (\partial \A)^4) \propto \op{Tr} ( \A (\partial \A)^4 ) \label{eqn_gs_SO(32)} 
\end{equation}
As is well-known, for $so(n)$, the trace in the vector and adjoint representation are related by 
\begin{equation*} 
	\op{Tr} (X^6) = (n-32) \op{tr} (X^6) + 15 (\op{tr}(X^2) ) (\op{tr}(X^4)).
\end{equation*}
Only when $n = 32$ does equation \eqref{eqn_gs_SO(32)} hold.

For the super Lie algebras $so(32+N \mid N)$, \eqref{eqn_gs_SO(32)} continues to hold. The main point is that the trace of a power of an element of $\mf{so}(32)$ in the adjoint representation of $\mf{so}(32+N \mid N)$ is the same as its trace in the adjoint of $\mf{so}(32)$. 

We conclude that the twisted type I string is consistent only when we include $(32+N \mid N)$ $D9$ branes, just as in the case of the physical string.

At higher orders in the loop expansion, the cancellation that occurs for the type I topological string on $\C^3$ holds similarly on $\C^5$.  The argument is identical to the one given before. First, we try to quantize the open-closed theory with gauge Lie algebras $\mf{so}(32+N \mid N)$ for all $N$, in a compatible way.  This means that we remove all trace relations from the open-string sector.     The single-trace operators of holomorphic Chern-Simons theory are given by the components $T_l^{(k,\bullet)}$ of the stress-energy tensor, where $k+l$ is odd and $l \ge 0$.  The operators from the closed-string sector gives us $T_l^{(k,\bullet)}$ when $l < 0$ and $k+l$ is odd. The BRST operator is as in equation \eqref{BRST-variation-coupled}, so that at the level of BRST cohomology only the topological operators $T_l^{(0,0)}$, $l$ odd, remain. Because these operators are of dimension $0$, they can not contribute to anomalies or counter-terms at higher orders in the loop expansion.

\noindent \textbf{Other gauge groups}

The Green-Schwarz mechanism \cite{GS} allows for the groups $E_8 \oplus E_8$, $E_8 \oplus U(1)^{248}$, $U(1)^{496}$ in addition to $SO(32)$.  String-theoretic reasoning selects $SO(32)$.

In our setting, the cancellation of one-loop anomalies works for any Lie algebra $\mf{g}$ in which the trace $\op{Tr}_{ad} X^6$ (taken in the adjoint representation) factors as a product of the Killing form and a quartic invariant polynomial. This happens for all the Lie algebras listed above, but also for some other algebras such as $\mf{su}(n)$ for $n \le 5$, $E_8$, etc. Unlike the standard Green-Schwarz calculation, our method does not immediately produce the condition that the Lie algebra is of dimension $496$.


In our analysis, we do not see the constraint that the dimension is $496$ directly. However, the argument for higher-order cancellation of open and closed anomalies \emph{only} works for $SO(32)$. For a gauge group such as $E_8$ or $E_8 \oplus E_8$, the coupled open-closed theory has non-topological local operators which can contribute to anomalies and counter-terms at higher orders in the loop expansion. To see this, we note that  the closed-string field in $\PV^{3,\ast}$ is no longer coupled to any open-string field.  This coupling is what induces the BRST differential cancelling open and closed sectors.  This tells us that any local operators we can build from the closed-string field in $\PV^{1,\ast}$ will survive when we couple to the gauge theory. 

To sum up, we see that although we do not directly find the constraint that the gauge group is of dimension $496$, our anomaly cancellation method only works for $OSp(32 + N\mid N)$, so ultimately we find the same gauge group that is selected by string theory.

\subsection{Branes in type I string theory} 
Physical type I string theory has $D1$ and $D5$ branes (as well as the $32$ $D9$ branes we have already discussed). In this section we will explain their appearance in the twisted type I string theory that we have described above. We will also describe the branes that appear in type I topological string theory in $3$ complex dimensions. 


Our general analysis of branes applies to complex dimensions $3$ or $5$.  In the ordinary topological $B$-model on a Calabi-Yau manifold $X$ of dimension $3$ or $5$, a brane is given by a coherent sheaf $\mc{F}$ on $X$.  The $\Z/2$ action which reverse orientation on the world-sheet acts on the category of branes by sending a vector bundle $E$ to the dual bundle $E^\vee$. A coherent sheaf $\mc{F}$ is sent to its dual in the derived sense,
\begin{equation} 
	\mc{F}^\vee = \underline{ \R\op{Hom} } (\mc{F}, \Oo_X ). 
\end{equation}
In practice, if we choose a complex of vector bundles as a resolution of $\mc{F}$, then $\mc{F}^\vee$ is computed by the dual of this complex. A brane in the orbifold theory is a sheaf $\mc{F}$ with a symmetric isomorphism $\mc{F} \iso \mc{F}^\vee$ (again, in the derived category).  Symmetric means that the pairing
\begin{equation} 
	\mc{F} \otimes^{\mbb L} \mc{F} \to \Oo_X 
\end{equation}
is symmetric.

If $Y \subset X$ is a complex submanifold of complex codimension $k$, then one can show that
\begin{equation} 
	\Oo_Y^\vee = \wedge^k N_{Y/X} [-k] 
\end{equation}
where $N_{Y/X}$ is the normal bundle to $Y$ in $X$ and $[-k]$ indicates a shift of cohomological degree by $k$. Since $X$ is Calabi-Yau, $\wedge^k N_{Y/X}$ is the same as the canonical bundle $K_Y$ of $Y$, so that
\begin{equation} 
	\Oo_Y^\vee = K_Y[-k]. 
\end{equation}
From this we see that we can build a self-dual sheaf wrapping $Y$ only if the codimension $k$ is even, and if we equip $Y$ with a square root $K_Y^{1/2}$ of the canonical bundle. In that case, the self-dual sheaf is $K_Y^{1/2} [-k/2]$, since
\begin{equation*} 
	(K_Y^{1/2}[-k/2])^\vee = K_Y^{-1/2}[k/2] \otimes \Oo_Y^\vee = K_Y^{1/2}[-k/2]. 
\end{equation*}

A self-dual sheaf $\mc{F}$ has, by definition, a map in the derived category $\mc{F} \otimes_{\Oo_X} \mc{F} \to \Oo_X$. We need this pairing to be symmetric. It can be shown that the pairing for $\mc{F}=K_Y^{1/2}[-k/2]$ obtained above 
is (graded) symmetric when $k/2$ is even, and (graded) anti-symmetric when $k/2$ is odd. 

When $k/2$ is even, we can tensor $K_Y^{1/2}[-k/2]$ with the fundamental representation $\C^n$ of $O(n,\C)$. The resulting sheaf $K_Y^{1/2}[-k/2] \otimes \C^n$ is still self-dual, and the pairing is symmetric.  We conclude that when $k/2$ is even, we can take a stack of $n$ branes for any $n$, and the theory on the brane is an $O(n)$ gauge theory.  

When $k/2$ is odd, the pairing on $K_Y^{1/2}[-k/2]$ is anti-symmetric. If we tensor with the fundamental representation $\C^{2n}$ of $Sp(n,\C)$, we obtain a sheaf $K_Y^{1/2}[-k/2] \otimes \C^{2n}$ which has a symmetric pairing.  We conclude that when $k/2$ is odd, we can only have an even number of branes, and that the gauge group on the brane is $Sp(n,\C)$.

For the type I $B$-model on a Calabi-Yau $5$-fold $X$, this argument tells us that the only branes are the $D1$ and $D5$ branes. We can have any number of $D1$ branes, and these have an $O(n)$ gauge group;  but only an even number of $D5$ branes, with an $Sp(n)$ gauge group.  This is consistent with what happens in physical type I \cite{Witten:1995gx}, where it is known that the gauge group on a stack of $D1$ branes is $O(n)$ and on a stack of $D5$ branes is $Sp(n)$.  

If we work on a Calabi-Yau $3$-fold, the only consistent brane in the type I theory is a $D1$ brane, living on an algebraic curve.  Since this is of complex codimension $2$, there must be an even number of $D1$ branes, carrying an $Sp(n)$ gauge group.  
\subsection{The open-string field theory on a brane in the type I topological string}

In this section  we will compute the open-string field theory living on a brane in the type I topological string, in dimensions $5$ or $3$.  In each case, there is potentially a one-loop gauge anomaly on the brane. However, we will find that $32$ (or $8$) is precisely the correct number of space-filling branes to cancel the anomaly on the brane (in dimensions $5$ and $3$). 

\noindent \textbf{The $D1$ brane in $5$ complex dimensions}

In the ordinary $B$-model on $\C^5$,  the open-string states for $n$ $D1$ branes are $\Omega^{0,\ast}(\C) [\eps_1,\dots,\eps_4] \otimes \gl(n)[1]$ where $\eps_i$ are odd variables. This is obtained from the dimension reduction of the usual holomorphic Chern-Simons theory on $\C^5$ to $\C=\{z_1=z_2=z_3=z_4=0\}$, with the identification $\epsilon_i= d\bar z_i$. 

We interpret these fields as follows: the $\eps_i$'s and the $\eps_i \eps_j \eps_k$'s are dual to each other and  give $4$ adjoint valued $\beta-\gamma$ systems, describing the normal motions of the brane.  The $\eps_i \eps_j$ fields give $6$ adjoint-valued real fermions (or $3$ adjoint-valued complex fermions).  The term without any $\eps$ gives the $\c$-ghost, and $\eps_1 \eps_2 \eps_3\eps_4$ is the $\b$-ghost.  Thus, the theory is the BRST reduction of $4$ $\beta-\gamma$ systems and $6$ real fermions, all adjoint-valued.  

The variables $\eps_i$ are most naturally of spin $1/4$, but the spin can be changed by twisting using a homomorphism $SO(2) \to SU(4)$.  If we choose the $\eps_i$ to have spin $1/4$, then the components of the $\beta-\gamma$ system are of spins $1/4$, $3/4$ and all other fields have their usual spin.

For type I, we should take the $\Z/2$ fixed points of these fields.  The $\Z/2$ action sends an element $A \in \mf{gl}(n)$ to $-A^T$, and also acts on the variables $\eps_i$ by $\eps_i \to -\eps_i$.
This extra sign is dictated by the requirement that the open-string fields act as symmetries of the sheaf of $K_{\C}^{1/2} \otimes \C^n[-2]$ which are graded anti-symmetric with respect to the pairing 
\begin{equation*} 
	(K_{\C^{1/2}} \otimes  \C^n)[-2] \otimes (K_{\C^{1/2}} \otimes  \C^n)[-2]\to \Oo_{\C^5}. 
\end{equation*}
We conclude that the $D1-D1$ strings in type I contribute an $O(n)$ $\b-\c$ ghost system coupled to $6$ real fermions valued in the adjoint representation and $4$ $\beta-\gamma$ systems valued in the symmetric square of the vector representation of $O(n)$\footnote{In the ordinary topological $B$-model, there is also a term in the BRST current of the form $f_{abc} \psi_a \beta_b \gamma_c$ where $a,b,c$ are indices for a basis of $\mf{gl}_N$.   When restricted to the fields of the type I $B$-model, this term vanishes for symmetry reasons.  }.   The symmetric square of the vector representation contains a copy of the trivial representation, so we can write the bosonic fields as being four free $\beta-\gamma$ systems, representing the normal motion, together with four $\beta-\gamma$ systems valued in the trace-free symmetric matrices.  

In addition, there is a contribution from the $32$ $D9$ branes. Because of the $\Z/2$ orbifold, only $D9-D1$ strings contribute; $D1-D9$ strings do not contribute.  For $n$ $D1$ branes, these fields live in 
\begin{equation} 
	\op{RHom}(\Oo_{\C^5}^{32}, K_{\C}^{1/2} \otimes \C^n[-2])[1] = K_{\C}^{1/2} \otimes \C^n \otimes \C^{32} [-1].    
\end{equation}
(The shift of ghost number by $1$ is the shift that always occurs when we move from open-string states to fields of the gauge theory).  Since only the ghost number modulo $2$ really matters, we see that these extra fields contribute $32n$ real fermions.

Note that the theory on the $D1$ branes is \emph{not} the dimensional reduction of the $SO(n)$ holomorphic Chern-Simons theory on a space-filling brane.  This is to be expected: there is no $T$-duality in type I string theory which transforms $D1$ to $D9$, nor is there a $T$-duality transforming $D1$ to $D5$.

The theory on the stack of $n$ $D1$ branes is anomaly-free.  Using the standard formula for the chiral anomaly to BRST reduction of a system of fermions and $\beta-\gamma$ systems, we find the anomaly is
\begin{equation} 
	8 \op{Tr}_{S^2 \C^n }(X^2) - 6 \op{Tr}_{\wedge^2 \C^n }(X^2) - 32 \op{Tr}_{\C^n}(X^2) - 2 \op{Tr}_{\wedge^2 \C^n}(X^2). \label{eqn:open-string_anomaly} 
\end{equation}
Here $X$ is an element of the Cartan of $\mf{so}(n)$, and $\C^n$ indicates the fundamental representation. The four terms come from the four $\beta-\gamma$ systems in $S^2 \C^n$, the $6$ adjoint valued fermions, the $32$ vector valued fermions, and the $\b-\c$ ghosts.  

One can calculate that this combination is zero. Indeed, we can take $X$ to be a generator of the Cartan of a copy of $\mf{so}(2) \subset \mf{so}(n)$, and break $\C^n = \C^2 \oplus \C^{n-2}$ as a representation of $\mf{so}(2)$.  Then, $S^2 \C^n = S^2 \C^2 \oplus \C^2 \otimes \C^{n-2} \oplus S^2 \C^{n-2}$, and $\wedge^2 \C^n = \C \oplus \C^2 \otimes \C^{n-2} \oplus \wedge^2 \C^n$. This gives
\begin{align*} 
	\op{Tr}_{S^2 \C^n} (X^2) &= \op{Tr}_{S^2 \C^2} (X^2) + (n-2) \op{Tr}_{\C^2}(X^2) = 8 + 2(n-2) \\
\op{Tr}_{\wedge^2 \C^n} (X^2) &=  (n-2) \op{Tr}_{\C^2}(X^2) =  2(n-2) \\
	\op{Tr}_{\C^n}(X^2) &= \op{Tr}_{\C^2}(X^2) = 2. 
\end{align*}
In these formulae, we are normalizing $X$ in the Cartan of $\mf{so}(2)$ so that the weights of the vector representation are $(1,-1)$.This implies that the anomaly \eqref{eqn:open-string_anomaly} vanishes.  Note that the vanishing of this anomaly gives an independent way of fixing the number of $D9$ branes to be $32$.

\noindent \textbf{$D1$ brane in three complex dimensions}
A similar argument gives a description of the theory living on a $D1$ brane in $3$ complex dimensions.  In this case, we find $\b-\c$ ghosts for the group $\mf{sp}(n)$. If $\C^{2n}$ is the fundamental representation of $\mf{sp}(n)$, then the matter fields consist of a  $\beta-\gamma$ system living in $\wedge^2 \C^{2n}$ together with a system of symplectic bosons living in the bifundamental representation $\C^8 \otimes \C^{2n}$ of $\mf{so}(8) \oplus \mf{sp}(n)$.   The trivial representation in $\wedge^2(\C^{2n})$ is a free $\beta-\gamma$ system describing the normal fluctuations of the brane.  

In this case also, there still is no anomaly to performing the BRST reduction. The anomaly consists of 
\begin{equation} 
	2 \op{Tr}_{\wedge^2 \C^{2n}} (X^2) + 8 \op{Tr}_{\C^{2n}} (X^2) - 2 \op{Tr}_{S^2 \C^{2n}}(X^2). 
\end{equation}
These terms come from the $\beta-\gamma$ system valued in $\wedge^2 \C^{2n}$, the $8$symplectic bosons, and the $\b-\c$ ghosts.  As before, $X$ is an element of the Cartan of $\mf{sp}(2n)$. 

To check this expression vanishes, we can assume that $X$ is a basis element the Cartan of $\mf{sp}(1)$ and we decompose $\C^{2n} = \C^2 \oplus \C^{2n-2}$, as above.  We find, following the analysis above, that $\op{Tr}_{\wedge^2 \C^{2n}} (X^2) = 2 (2n-2)$, $\op{Tr}_{S^2 \C^{2n}} (X^2) = 8 + 2(2n-2)$, $\op{Tr}_{\C^{2n}}(X^2) = 2$.   The total anomaly is then zero.  

This anomaly only vanishes if we have $8$ space-filling branes, giving us an independent verification that this is the correct number.

\noindent \textbf{The $D5$ brane in $5$ complex dimensions}
The theory on a stack of $2n$ $D5$ brane in $5$ complex dimensions can be calculated using similar techniques.  To describe this theory it is convenient to work in the BV formalism.  We will not present the full details of the derivation, as they are similar to the computations above.

The fundamental fields are:
\begin{equation} 
	\begin{split}
		\mc{A} & \in \Omega^{0,\ast}(\C^3) \otimes \mf{sp}(n) [1] \\
		\mc{B} & \in \Omega^{3,\ast}(\C^3) \otimes \mf{sp}(n) [1] \\
		\mc{X} & \in \Omega^{0,\ast}(\C^3, K^{1/2}) \otimes \wedge^2 \C^{2n} \\
		\mc{Y} & \in \Omega^{0,\ast}(\C^3, K^{1/2}) \otimes \wedge^2 \C^{2n} \\
		\mc{Z} & \in \Omega^{0,\ast}(\C^3, K^{1/2}) \otimes \C^{2n} \otimes \C^{32} 
	\end{split}
\end{equation}
The Lagrangian is
\begin{equation} 
	\int \mc{B} F(\mc{A}) + \mc{X} \dbar_{\mc{A}} \mc{Y} + \mc{Z} \dbar_{\mc{A}} \mc{Z}   
\end{equation}
The fields $\mc{X}, \mc{Y}, \mc{Z}$ are those of holomorphic Rozansky-Witten theory valued in the symplectic manifold $\C^{2n} \otimes \C^{32} \oplus \wedge^2 \C^{2n} \otimes \C^2$.  The fields $\mc{A}$, $\mc{B}$ gauge holomorphic Rozansky-Witten theory by coupling to holomorphic BF theory for the gauge group $\mf{sp}(n)$. 

There is a one-loop anomaly given by the same diagram which leads to the one-loop anomaly for holomorphic Chern-Simons theory.  The expression for the anomaly is
\begin{equation} 
	2 \op{Tr}_{\wedge^2 \C^{2n}} (\mc{A} (\partial \mc{A})^3 ) +  32 \op{Tr}_{\C^{2n}} (\mc{A} (\partial \mc{A})^3 ) -  2 \op{Tr}_{S^2 \C^{2n}} (\mc{A} (\partial \mc{A})^3 ) . 
\end{equation}

Remarkably enough, $32$ is precisely the correct number of $D9$ branes to make this anomaly cancel. To see this we need the Lie algebraic identity:
\begin{equation} 
	2 \op{Tr}_{\wedge^2 \C^{2n}} (X^4 ) +  32 \op{Tr}_{\C^{2n}}  (X^4 )=  2 \op{Tr}_{S^2 \C^{2n}} (X^4 ) \label{eqn:sp_identity}
\end{equation}
for $X$ in the Cartan of $\mf{sp}(n)$.  To verify this, we can assume that $X$ is in the Cartan of $\mf{sp}(1)$, and decompose $\C^{2n} = \C^2 \oplus \C^{2n-2}$ as an $\mf{sp}(1)$ representation.  Then $\wedge^2 \C^{2n} = \C \oplus \C^2 \otimes \C^{2n-2} \oplus \wedge^2 \C^{2n-2}$ and $S^2 \C^{2n} = S^2 \C^2 \oplus \C^2 \otimes \C^{2n-2} \oplus S^2 \C^{2n-2}$ as $\mf{sp}(1)$ representations.  For $X$ a basis of the Cartan of $\mf{sp}(1)$  we have $\op{Tr}_{\C^2} X^4  = 2$ and $\op{Tr}_{S^2 \C^2} X^4 = 32$. This implies equation \eqref{eqn:sp_identity}.

\subsection{The $D1$ brane and the heterotic string}
Under the duality between type I and the $\op{Spin}(32)/Z_2$ heterotic string \cite{W3,PW} the theory on the $D1$ brane becomes the world-sheet theory of the heterotic string.  As a final consistency check, we will show that in our twisted version of type I string theory, the theory on the $D1$ brane is a supersymmetric localization of the heterotic string world-sheet theory.  

Let us consider the theory living on a single $D1$ brane in the type I $B$-model on $\C^5$.  Since $SO(1)$ is trivial, the $\b-\c$ ghost system and the $6$ adjoint valued fermions do not contribute.  The gauge group on the $D1$ brane lives in $O(1)$, however, so we have an extra $\Z/2$ gauge symmetry which will be crucial for the analysis. 

The heterotic string can be described \cite{Gr} as $32$ real fermions coupled to the $\sigma$-model with $(0,2)$ supersymmetry with target the $8$ transverse directions to the worldsheet.  There is a slightly subtle procedure for restricting to only certain states of the system of fermions, which differs between the $\op{Spin}(32)/\Z_2$ and $E_8 \times E_8$ heterotic strings.  We will find precisely the fermions of the $\op{Spin}(32)/\Z_2$ model.

First let us describe the appearance of the $(0,2)$ system.  It is shown in \cite{W4} that a $(0,2)$ $\sigma$-model can be twisted to yield the $\beta-\gamma$ system.  The $D1-D1$ strings in our model, in the case $n= 1$, yield a $\beta-\gamma$ system on $\C^4$, the four complex transverse directions to the worldsheet.  This matches the twist of this part of the heterotic string world-sheet theory.

The $D9-D1$ strings yield, as we have seen, $32$ real fermions.  These live, of course, in $K_{\C}^{1/2}$.  The choice of spin structure on the world-sheet is part of the data of a self-dual sheaf on $\C^5$. As such, it must be dynamical. This tells us that both Ramond-Ramond and Neveu-Schwarz sectors of the fermions must contribute.   
In addition, the $O(1) = \Z/2$ gauge group of the world-sheet theory acts in a non-trivial way on the system of fermions, and cuts down the physical states. In the Neveu-Schwarz sector, there are no zero-modes, and so the space of states is obtained by applying the modes $\psi^i_{n/2}$, $n > 0$, $i = 1,\dots,32$ of the fermion to the vacuum vector.  The action of $\Z/2$ sends $\psi^i \to -\psi^i$, so that when we pass to gauge invariants only states involving an even number of fermions contribute.  This matches the constraint in \cite{Gr}.

In the Ramond sector, the zero modes form the Clifford algebra $\op{Cl}_{32}$.  The space of states is then generated by applying the operators $\psi^i_n$, $n > 0$ to the irreducible representation of the Clifford algebra. If $S_{\pm}$ denote the two irreducible spin representations of $SO(32)$, then the Clifford algebra representation is $S_+ \oplus S_-$.  The prescription of \cite{Gr} is that we should take the ground states of the physical Hilbert space to be given those elements of $S_+ \oplus S_-$ on which the operator
\begin{equation*} 
	 \gamma =\psi^1_0 \dots \psi^{32}_0  
\end{equation*}
acts as the identity. 

In our story, this prescription is achieved as follows.  The action of $\Z/2$ on the fermions, sending $\psi^i_n \to -\psi^i_n$, acts on the operators in the Ramond sector.  We need to lift this to a compatible action on the Hilbert space of the Ramond sector.  The Hilbert space is freely generated over the spin representation $S_+ \oplus S_-$ by $\psi^i_{n}$ for $n > 0$.  Once we define the $\Z/2$ action on $S_+ \oplus S_-$ we get it on the whole space of states.

To define the action of $\Z/2$ on the Clifford module, we need to make the $\Z/2$ action on the Clifford algebra into an inner action. That is, we need to construct some operator $\eta$ such that $\eta^2 = 1$ and $\eta \psi^i_0 \eta = -\psi^i_0$.  Setting $\eta = \gamma$ as above satisfies these properties. This operator acts by $1$ on $S_+$, and $-1$ on $S_-$.  

We conclude that the Hilbert space in the Ramond sector consists of states generated from an element of $S_+$ by an even number of $\psi^i_n$, and from $S_-$ by an odd number of $\psi^i_n$.   This matches the description of \cite{Gr}. 

In sum, we have found that the theory on a single $D1$ brane in our proposed twist of type I string theory matches \emph{precisely} with a supersymmetric localization of the $\op{Spin}(32)/\Z_2$ heterotic string.  Given the known duality between type I and heterotic strings, this result gives convincing evidence as that type I Kodaira-Spencer theory, coupled to $SO(32)$ holomorphic Chern-Simons theory, is indeed a supersymmetric localization of the space-time theory of the type I superstring.

\begin{acknowledgments}

The work of K.C. is supported by the NSERC Discovery Grant program and by the Perimeter Institute for Theoretical Physics. Research at Perimeter Institute is supported by the
Government of Canada through Industry Canada and by the Province of
Ontario through the Ministry of Research and Innovation.

The work of S.L. is partially supported by grant 11801300 of National Natural Science Foundation of China  and grant Z180003 of Beijing Municipal Natural Science Foundation. Part of this work was done when S.L. was visiting Perimeter Institute for Theoretical Physics in Jan 2019. S.L. thanks for their hospitality and provision of excellent working enviroment.

\end{acknowledgments}

\end{document}